\documentclass[twocolumn,showpacs,preprintnumbers,amsmath,amssymb,pre]{revtex4}

\usepackage{graphicx}
\usepackage{dcolumn}
\usepackage{bm}
\usepackage{enumerate}
\usepackage{amsmath}
\usepackage{color}

\begin{document}

\title{Cattaneo--type subdiffusion equation}

\author{Tadeusz Koszto{\l}owicz}
 \email{tadeusz.kosztolowicz@ujk.edu.pl}
 \affiliation{Institute of Physics, Jan Kochanowski University,  Uniwersytecka 7, 25-406 Kielce, Poland}
\affiliation{Department of Radiological Informatics and Statistics, Medical University of Gda\'nsk, Tuwima 15, 80-210 Gda\'nsk, Poland}

 \author{Aldona Dutkiewicz}
 \email{szukala@amu.edu.pl}
 \affiliation{Faculty of Mathematics and Computer Science, Adam Mickiewicz University,
 Uniwersytetu Pozna\'nskiego 4, 61-614 Pozna\'n, Poland}

\author{Katarzyna D. Lewandowska}
 \email{katarzyna.lewandowska@gumed.edu.pl}
 \affiliation{Department of Physics and Biophysics, Medical University of Gda\'nsk,
 D\c{e}binki 1, 80-211 Gda\'nsk, Poland}

\date{\today}

\begin{abstract}
The ordinary subdiffusion equation, with a fractional time derivative of at most first order, describes a process in which the propagation velocity of diffusing molecules is unlimited. To avoid this non-physical property different forms of the Cattaneo subdiffusion equation have been proposed. 
We define the Cattaneo effect as a delay of the ordinary subdiffusion flux activation by a random time. 
By incorporating this effect into the flux equation we get a Cattaneo--type subdiffusion equation (CTSE).
We study the CTSE that differs from the ordinary subdiffusion equation by an additional integro--differential operator (AO) controlled by a time delay probability distribution. A method for deriving CTSE within the standard continuous time random walk model is also shown.
As examples, we consider the CTSE with AO being the Caputo fractional time derivative of the order independent of the subdiffusion exponent and with the AO with a kernel that is a slowly varying function. In the first case the Cattaneo effect disappears over time much faster than in the second one. 
Based on the Green's functions, the time evolutions of the first passage time distribution and of the mean square displacement of a diffusing molecule, we discuss whether the influence of the Cattaneo effect is significant. In the considered examples, this influence seems to be small. However, a relative probability of finding a molecule at a long distance from the starting point for the CTSE equation with respect to the ordinary subdiffusion equation increases rapidly with distance. Even small changes caused by the Cattaneo effect can lead to different results in modeling processes where a faster appearance of a diffusing object changes the nature of the process. The effect may be important, for example, in modeling the spread of an epidemic when a diffusing object is a source of infection.
\end{abstract}

\maketitle

\section{Introduction\label{secI}}

Subdiffusion occurs in media in which particle random walk is very hindered. The examples are transport of molecules in viscoelastic chromatin network \cite{lee2021}, in porous media \cite{bijeljic2011}, and in living cells \cite{barkai2012}, transport of sugars in agarose gel \cite{kdm}, and antibiotics in bacterial biofilm \cite{km}. 

In normal diffusion, the mean square displacement (MSD) $\sigma^2$ of a diffusing molecule grows linearly with time, $\sigma^2(t)\sim t$. In general, subdiffusion is a process in which $\sigma^2$ evolves more slowly than it does for normal diffusion. In ordinary subdiffusion there is $\sigma^2(t)=2Dt^\alpha/\Gamma(1+\alpha)$, where $\alpha\in(0,1)$ is the subdiffusion parameter (exponent), $D$ is a subdiffusion coefficient given in units of ${\rm m^2/s^\alpha}$, while for slow subdiffusion (ultraslow diffusion) $\sigma^2$ is controlled by a slowly varying function, such as logarithm. We consider subdiffusion in a one-dimensional homogeneous system with constant parameters describing the process.

Frequently, ordinary subdiffusion is described by a fractional differential equation \cite{wyss1986,hilferanton,compte,mks,mk,barkai2000,skb,klages2008,ks} with the Riemann-Liouville fractional time derivative of the order $1-\alpha$ or an equation with the Caputo one of the order $\alpha$. 
For the parabolic normal diffusion equation and subdiffusion equation with a time derivative of at most first order the probability density of finding the molecule at position $x$ at time $t$ (the Green's function) $P(x,t|x_0)$, where $x_0$ is the initial particle position, is greater than zero for any $x$ and $t>0$. It means that the propagation velocity of some particles is arbitrarily high. To avoid this non--physical property, the Cattaneo normal diffusion equation has been proposed; the equation was originally used to describe heat propagation \cite{cattaneo}. The equation involves a second-order time derivative controlled by the parameter $\tau$, see Eq. (\ref{eq15}) presented later. One of the models that provides the equation is the phase lag model, where the flux of diffusing particles is a temporal convolution of a decreasing exponential function and a particles concentration gradient \cite{zhmakin}.
 
While the Cattaneo hyperbolic normal diffusion equation is well defined, the Cattaneo subdiffusion equation can take various non-equivalent forms, see Refs. \cite{compte1997,fernandez,qi,liu,roscani,alegria,luchko,awad2021,awad2019,hamada,metzler1999,metzler1998,gorska2020,gorska2021,koszt2014}. The Cattaneo equation with the Caputo and/or the Riemann--Liouville fractional derivatives have been used to describe subdiffusion of neutrons inside the core of a nuclear reactor \cite{vya}, heat transport in porous media \cite{nikan}, and in a system with glass spheres in a tank filled with air \cite{moza}. Cattaneo--type equations with fractional derivatives other than those mentioned above were also considered. The examples are the equations with Caputo--Fabrizio fractional time derivative \cite{liu2017}, with tempered Caputo derivative \cite{beghin}, and with Hilfer fractional derivative with respect to another function \cite{vieira}. 

We define the Cattaneo effect as a delay of the activation of ordinary subdiffusion flux by a random time. Combining the constitutive equation, which describes the relation between the probability density flux and the gradient of Green's function, with the continuity equation the Cattaneo--type subdiffusion equation (CTSE) is obtained. We pay special attention to the CTSE that differs from the ordinary subdiffusion equation by an additional term with integral operator acting on the time variable. The kernel of this operator is controlled by a probability density of delay time. The additional term describes the Cattaneo effect that change ordinary subdiffusion. 

The time evolution of MSD does not clearly determine the type of diffusion \cite{dybiec1,meroz2011}. As shown in Ref. \cite{dybiec1}, the appropriate combination of superdiffusion (i.e. facilitated diffusion for which $\sigma^2(t)\sim t^\nu$ with $\nu>1$) and subdiffusion effects leads to a process in which $\sigma^2(t)\sim t$, but such a process is certainly not normal diffusion. In addition to MSD, we also consider the first passage time distribution $F(t;x)$ of a particle through an arbitrarily chosen point $x$.

If the diffusion properties of the system do not change over time, the type of diffusion, as well as parameters describing the process, should not change either (assuming that the parameters do not depend on the concentration of diffusing substance). However, mathematical properties of solutions to the Cattaneo-type subdiffusion equation are in some contradiction with the above-mentioned physical properties.
When diffusion in a subdiffusive medium is described by the Cattaneo--type equation, for very short time there is $\sigma^2(t)\sim t^\nu$ with $\nu>1$. For example, there are $\nu=2\alpha$, $\nu=1+\alpha$, and $\nu=2$ for different models considered in Ref. \cite{compte1997}. The appearance of a term with an exponent greater than 1 is typical for superdiffusion. The relation $\sigma^2(t)\sim t^\alpha$ with $\alpha\in(0,1)$, typical for subdiffusion, appears for longer times. It seems to be unphysical that superdiffusion could occur in a subdiffusive medium whose diffusion properties do not change over time. Therefore, in the following we consider the Cattaneo-type subdiffusion equation for time at which the process can be interpreted as subdiffusion.

Solutions to the Cattaneo--type subdiffusion equation and solutions to the subdiffusion equation without the Cattaneo effect are usually not much different from each other. Then, the latter equation, which appears to be easier to solve, is used to describe subdiffusion. However, there are processes that both equations provide qualitatively different results even for small Cattaneo effect. Example of this is the diffusive and subdiffusive impedance \cite{barbero,kostrobij,kl2009,lk2008}. When considering subdiffusion described by a Cattaneo-type equation, the question arises whether such an equation gives a significant difference compared to the use of the ordinary subdiffusion equation. We show that although the differences in the functions describing subdiffusion obtained from the ordinary subdiffusion equation and CTSE are quite small, the relative differences of these functions can be quite large. The latter property suggests that the Cattaneo effect may significantly change the model. This problem will be briefly discussed in Sec. \ref{secVIII}.

The article is organized as follows. The general form of Cattaneo--type subdiffusion equation is discussed in Sec. \ref{secII}. The probability distribution of the delay time $R$ controls the Cattaneo effect. In Sec. \ref{secIII} the  functions describing subdiffusion, namely Green's function, first passage time distribution, and time evolution of mean square displacement of diffusing particle, are described. The explicit forms of the functions are given in terms of the Laplace transform. In Sec. \ref{secIV}, we consider the probability distribution $R$ providing the CTSE which can be interpreted as an ordinary subdiffusion equation with an additional operator (AO) added; AO depends on the Cattaneo effect. Special cases of CTSE with the AO being the Caputo fractional time derivative of the order independent of the subdiffusion exponent and CTSE with the AO generated by a slowly varying function are studied in Secs. \ref{secV} and \ref{secVI}, respectively. We pay attention how the Cattaneo effect affects ordinary subdiffusion. A method for deriving CTSE within the standard continuous time random walk model is shown in Sec. \ref{secVII}. Final remarks are presented in Sec. \ref{secVIII}.

\section{Subdiffusion equation with Cattaneo effect\label{secII}}

In the following, we will denote functions describing a process in which the Cattaneo effect does not occur by $\tau=0$, the interpretation of the parameter $\tau$ will be presented later in this section and in Sec. \ref{secIV}.
The ordinary subdiffusion equation, derived within the continuous time random walk (CTRW) model \cite{mk,ks}, reads
\begin{equation}\label{eq1}
\frac{\partial P_{\tau=0}(x,t|x_0)}{\partial t}=D\frac{^{RL}\partial^{1-\alpha}}{\partial t^{1-\alpha}}\frac{\partial^2 P_{\tau=0}(x,t|x_0)}{\partial x^2},
\end{equation}
where $\alpha\in(0,1)$. The Riemann--Liouville fractional derivative, involved in Eq. (\ref{eq1}), is defined as
\begin{equation}\label{eq2}
\frac{^{RL}d^\nu f(t)}{dt^\nu}=\frac{1}{\Gamma(n-\nu)}\frac{d^n}{dt^n}\int_0^t (t-u)^{n-\nu-1}f(u)du,
\end{equation}
where $\nu>0$ and $n$ is a natural number, $n=\lfloor\nu\rfloor+1$ when $\nu\notin\mathcal{N}$, otherwise $n=\nu\in\mathcal{N}$. 

When a lot of molecules diffuse independently of each other, their concentration $C$ can be calculated using the formula $C(x,t)=\int_{-\infty}^\infty C(x_0,0)P(x,t|x_0)dx_0$. Then, Eq. (\ref{eq1}) is also satisfied by the function $C(x,t)$. In the following we assume that the initial particle position at $t=0$ is $x_0=0$. In this case, to shorten the notation, we omit $x_0$ in the function notation, e.g. $P(x,t)\equiv P(x,t|0)$.

A diffusion equation can be obtained phenomenologically by a combination of the continuity equation and a constitutive equation defining the flux $J$ of diffusing molecules (or probability density flux when random walk of a single molecule is considered). The continuity equation reads 
\begin{equation}\label{eq3}
\frac{\partial P(x,t)}{\partial t}=-\frac{\partial J[P(x,t)]}{\partial x}.
\end{equation}
where $J[P]$ is the flux operator acting on the function $P$.  
For ordinary subdiffusion the constitutive flux equation is defined as
\begin{equation}\label{eq4}
J_{\tau=0}[P_{\tau=0}(x,t)]=-D\frac{^{RL}\partial^{1-\alpha}}{\partial t^{1-\alpha}}\frac{\partial P_{\tau=0}(x,t)}{\partial x}.
\end{equation}
 
The Cattaneo effect is defined as the ordinary subdiffusion flux $J_{\tau=0}$ activation delay by a random time, 
\begin{equation}\label{eq5} 
J[P(x,t)]=\int_0^t R(t';\tau)J_{\tau=0}[P(x,t-t')]dt',
\end{equation}
where $R$ is a probability density of the delay time. The Cattaneo--type subdiffusion equation is obtained by combination of Eqs. (\ref{eq3}) and (\ref{eq5}). 

When analyzing diffusion equations it is convenient to use the Laplace transform (LT) $\mathcal{L}[f(t)](s)=\int_0^\infty {\rm e}^{-st}f(t)dt\equiv \hat{f}(s)$. 
The LT of Eqs. (\ref{eq3}) and (\ref{eq5}) are, respectively,
\begin{equation}\label{eq6}
s\hat{P}(x,s)-P(x,0)=-\frac{\partial \hat{J}[\hat{P}(x,s)]}{\partial x},
\end{equation}
\begin{equation}\label{eq7}
\hat{J}[\hat{P}(x,s)]=\hat{R}(s;\tau)\hat{J}_{\tau=0}[\hat{P}(x,s)].
\end{equation}
The diffusion type is defined by the flux operator $J_{\tau=0}$, while the Cattaneo effect is controlled by the probability distribution $R$ which depends on the parameter $\tau$. We assume that the Cattaneo effect is turned off when $\tau=0$.

Since $R$ is a probability density of delay time, this function satisfies the following conditions: 
\begin{enumerate}
	\item $R(t;\tau)$ is normalized, $\int_0^\infty R(t;\tau)dt=1\Leftrightarrow\hat{R}(0;\tau)=1$,
	\item $R(t;\tau)$ is non-negative in the entire time domain. According to the Bernstein theorem, this property is satisfied when $\hat{R}(s;\tau)$ is a completely monotonic function ($\mathcal{CMF}$), i.e. it satisfies the condition $(-1)^n \partial^n\hat{R}(s;\tau)/\partial s^n\geq 0$ for $n=0,1,2,\ldots$ \cite{gorska2020,schilling}. In the following, we assume that $1/\hat{R}$ is a complete Bernstein function ($\mathcal{CBF})$, which makes $\hat{R}\in\mathcal{CMF}$, see Appendix A, property 11.
	\item the Cattaneo effect disappears when $\tau=0$, then $R(t;0)=\delta(t)\Leftrightarrow\hat{R}(s;0)=1$, where $\delta$ is the delta--Dirac function.  
\end{enumerate}

Due to the relation \cite{ks}, p.82,
\begin{equation}\label{eq8}
\mathcal{L}\Bigg[\frac{^{RL}d^\alpha f(t)}{dt^\alpha}\Bigg](s)=s^\alpha \mathcal{L}\left[f(t)\right](s),\;\alpha\in(0,1),
\end{equation}
the Laplace transforms of Eq. (\ref{eq4}) is
\begin{equation}\label{eq9}
\hat{J}_{\tau=0}[\hat{P}_{\tau=0}(x,s)]=-Ds^{1-\alpha}\frac{\partial\hat{P}_{\tau=0}(x,s)}{\partial x}.
\end{equation}
Combining Eqs. (\ref{eq6}), (\ref{eq7}), and (\ref{eq9}) we obtain
\begin{equation}\label{eq10}
s\hat{P}(x,s)-P(x,0)=D\hat{R}(s;\tau)s^{1-\alpha}\frac{\partial^2 \hat{P}(x,s)}{\partial x^2}.
\end{equation}
In the time domain Eq. (\ref{eq10}) reads
\begin{equation}\label{eq11}
\mathcal{B}_{t;\tau}[P(x,t)]
=D\frac{^{RL}\partial^{1-\alpha}}{\partial t^{1-\alpha}}\frac{\partial^2 P(x,t)}{\partial x^2},
\end{equation}
where
\begin{equation}\label{eq12}
\mathcal{B}_{t;\tau}[f(t)]=\int_0^t \mathcal{L}^{-1}\left[\frac{1}{\hat{R}(s;\tau)}\right](t')f^{(1)}(t-t')dt',
\end{equation}
$f^{(1)}(t)=df(t)/dt$.

We note that the normal diffusion equation and the normal diffusion flux can be formally obtained by putting $\alpha=1$ in the above equations. The Cattaneo effect for the normal diffusion equation is generated by the following probability distribution \cite{compte1997}
\begin{equation}\label{eq13}
R(t';\tau)=\frac{1}{\tau}{\rm e}^{-\frac{t'}{\tau}}.
\end{equation}
The Laplace transform of Eq. (\ref{eq13}) is
\begin{equation}\label{eq14}
\hat{R}(s;\tau)=\frac{1}{1+\tau s}.
\end{equation}

Combining Eqs. (\ref{eq10}) for $\alpha=1$ and (\ref{eq14}), and calculating the inverse Laplace transform (additionally assuming $P^{(1)}(x,0)=0$), the well--known Cattaneo normal diffusion equation is obtained
\begin{equation}\label{eq15}
\tau\frac{\partial^2 P(x,t)}{\partial t^2}+\frac{\partial P(x,t)}{\partial t}=D\frac{\partial^2 P(x,t)}{\partial x^2}.
\end{equation}

\section{Functions describing diffusion\label{secIII}}

The Green's function is the solution to diffusion equation with the boundary conditions
\begin{equation}\label{eq16}
P(\pm\infty,t)=0,
\end{equation} 
and the initial condition
\begin{equation}\label{eq17}
P(x,0)=\delta(x).
\end{equation}

In terms of the Laplace transform the Green's function for Eq. (\ref{eq11}) is 
\begin{equation}\label{eq18}
\hat{P}(x,s)=\frac{1}{2\sqrt{D\hat{R}(s;\tau)}s^{1-\alpha/2}}{\rm e}^{-\frac{s^{\alpha/2}|x|}{\sqrt{D\hat{R}(s;\tau)}}}.
\end{equation}
Eq. (\ref{eq18}) takes the form of Eq. (13) in Ref. \cite{gorska2020}. 

Since $P(x,t)$ is a probability density with respect to $x$ variable, this function must be normalized and non-negative. Integrating Eq. (\ref{eq18}) we get $\int_{-\infty}^\infty \hat{P}(x,s)dx=1/s$, thus the normalization condition is met. The results presented in Ref. \cite{gorska2020} provide that $P$ is non--negative when $\sqrt{s^\alpha/\hat{R}(s;\tau)}$ is a complete Bernstein function ($\mathcal{CBF}$). The proof of this statement uses conclusion from the Bernstein theorem that the function $P(x,t)$ is non-negative when $\hat{P}(x,s)$ is a completely monotone function ($\mathcal{CMF}$). The definitions and properties of $\mathcal{CMF}$ and $\mathcal{CBF}$ are presented in Refs. \cite{gorska2020,schilling} and in a shortened version in Appendix A in this paper. Let us assume that $1/\hat{R}\in\mathcal{CBF}$. Due to property 11 from Appendix A, this assumption provides $\hat{R}\in\mathcal{CMF}$ which is consistent with the condition 2 in Sec. \ref{secII}.  Thus, $s^{\alpha/2}/\sqrt{\hat{R}}\in\mathcal{CBF}$ (properties 6 and 10), $s\sqrt{\hat{R}}/s^{\alpha/2}\in\mathcal{CBF}$ (property 7), and $1/(\sqrt{\hat{R}}s^{1-\alpha/2})\in\mathcal{CMF}$ (property 11). Finally, properties 2 and 8 give $\hat{P}\in\mathcal{CMF}$.  

The Green's function Eq. (\ref{eq18}) is symmetrical, $P(x,t)=P(-x,t)$, this property is used in deriving further equations. A frequently used function that characterizes diffusion is the temporal evolution of the mean square displacement $\sigma^2$ of diffusing particle,
\begin{equation}\label{eq19}
\sigma^2(t)=2\int_0^\infty x^2 P(x,t)dx.
\end{equation}
From Eq. (\ref{eq18}) and the Laplace transform of Eq. (\ref{eq19}) we get
\begin{equation}\label{eq20}
\hat{\sigma^2}(s)=\frac{2D\hat{R}(s;\tau)}{s^{1+\alpha}}.
\end{equation}

Another function that characterizes diffusion is the first passage time distribution $F$. Assuming $x>0$, the distribution of time of the first particle passing the point $x$ can be calculated using the formula 
\begin{equation}\label{eq21}
F(t;x)=-\frac{\partial}{\partial t}\int_{-x}^x P(x',t)dx',
\end{equation}
the derivation of the above equation is in Appendix B.
In terms of the Laplace transform the above equation reads
\begin{equation}\label{eq22}
\hat{F}(s;x)=1-s\int_{-x}^x \hat{P}(x',s)dx'.
\end{equation}
Putting Eq. (\ref{eq18}) to Eq. (\ref{eq22}) we get
\begin{equation}\label{eq23}
\hat{F}(s;x)={\rm e}^{-\frac{xs^{\alpha/2}}{\sqrt{D\hat{R}(s;\tau)}}}.
\end{equation}

The continuity equation shows that the number of diffusing molecules in the system is constant. 
Assuming $J[P(-\infty,t)]=0$, integrating Eq. (\ref{eq3}) with respect to the $x$ variable, and calculating the time derivative, we obtain
\begin{equation}\label{eq24}
J[P(x,t)]=-\frac{\partial}{\partial t}\int_{-\infty}^x P(x',t)dx'.
\end{equation}
The Laplace transform of Eq. (\ref{eq24}) is
\begin{equation}\label{eq25}
\hat{J}[\hat{P}(x,s)]=1-s\int_{-\infty}^x \hat{P}(x',s)dx'.
\end{equation}
For $x>0$ we get
\begin{equation}\label{eq26}
\hat{J}[\hat{P}(x,s)]=\frac{1}{2}{\rm e}^{-\frac{xs^{\alpha/2}}{\sqrt{D\hat{R}(s;\tau)}}}.
\end{equation}
Eqs. (\ref{eq23}) and (\ref{eq26}) provide the relation
\begin{equation}\label{eq27}
J[P(x,t)]=\frac{1}{2}F(t;x),\;x>0.
\end{equation}
For $x<0$ the flux is negative, then 
\begin{equation}\label{eq28}
J[P(x,t)]=-\frac{1}{2}F(t;x),\;x<0.
\end{equation}

The question arises whether the Cattaneo subdiffusion equation brings a new quality in comparison with the ordinary subdiffusion equation. We compare the functions $F$ and $\sigma^2$ derived for $\tau=0$ and for $\tau\neq 0$. The functions $F_R$ and $\sigma^2_R$ show the time evolution of the relative change of first passage time distribution and MSD, respectively, 
\begin{equation}\label{eq29}
F_R(t;x)=\frac{F_{\tau=0}(t;x)-F(t;x)}{F_{\tau=0}(t;x)},
\end{equation}
\begin{equation}\label{eq30}
\sigma^2_R(t)=\frac{\sigma^2_{\tau=0}(t)-\sigma^2(t)}{\sigma^2_{\tau=0}(t)}.
\end{equation}
Additionally, we consider the relative change of the Green's function
\begin{equation}\label{eq31}
P_R(x,t)=\frac{P_{\tau=0}(x,t)-P(x,t)}{P_{\tau=0}(x,t)}.
\end{equation}

\section{Subdiffusion equation with an additional operator controlling the Cattaneo effect\label{secIV}}

Let us assume
\begin{equation}\label{eq32}
\hat{R}(s;\tau)=\frac{1}{1+\tau\hat{\gamma}(s)},
\end{equation}
where the parameter $\tau$ is given in units that make $\tau\hat{\gamma}(s)$ dimensionless.
Then, similarly to Eq. (\ref{eq15}), the Cattaneo effect is expressed by an additional operator (AO) added, as a separate term, to Eq. (\ref{eq1}). Due to properties 3, 4, and 9 from Appendix A we conclude that $\hat{\gamma}\in\mathcal{CBF}\Rightarrow (1/\hat{R})\in\mathcal{CBF}$. Thus, to ensure the non-negativity of the Green's function it is enough to assume that $\hat{\gamma}$ is a complete Bernstein function. The condition $\hat{\gamma}(0)=0$ gives the normalization of the function $R$. 

Eqs. (\ref{eq10}) and (\ref{eq32}) provide the Cattaneo--type diffusion equation written in terms of the Laplace transform,
\begin{equation}\label{eq33}
(\tau\hat{\gamma}(s)+1)[s\hat{P}(x,s)-P(x,0)]=Ds^{1-\alpha}\frac{\partial^2\hat{P}(x,s)}{\partial x^2}.
\end{equation}
In the time domain Eq. (\ref{eq33}) reads
\begin{equation}\label{eq34}
\tau\mathcal{D}[P(x,t)]+\frac{\partial P(x,t)}{\partial t}=D\frac{^{RL}\partial^{1-\alpha}}{\partial t^{1-\alpha}}\frac{\partial^2 P(x,t)}{\partial x^2},
\end{equation}
where 
\begin{equation}\label{eq35}
\mathcal{D}[f(t)]=\int_0^t \gamma(t')f^{(1)}(t-t')dt'.
\end{equation}

The Laplace transforms of Green's function and first passage time distribution are, respectively,
\begin{equation}\label{eq36}
\hat{P}(x,s)=\frac{\sqrt{1+\tau\hat{\gamma}(s)}}{2\sqrt{D}s^{1-\alpha/2}}{\rm e}^{-\frac{s^{\alpha/2}|x|\sqrt{1+\tau\hat{\gamma}(s)}}{\sqrt{D}}},
\end{equation}
\begin{equation}\label{eq37}
\hat{F}(s;x)={\rm e}^{-\frac{xs^{\alpha/2}\sqrt{1+\tau\hat{\gamma}(s)}}{\sqrt{D}}}.
\end{equation}
Assuming $\tau\hat{\gamma}(s)<1$, the power series with respect to $\tau$ of the above functions are
\begin{equation}\label{eq38}
\hat{P}(x,s)=\frac{1}{2\sqrt{D}s^{1-\alpha/2}}{\rm e}^{-\frac{|x|s^{\alpha/2}}{\sqrt{D}}}\sum_{i=0}^\infty \tau^i a_i(x,s)\hat{\gamma}^i(s),
\end{equation}
\begin{equation}\label{eq39}
\hat{F}(s;x)={\rm e}^{-\frac{|x|s^{\alpha/2}}{\sqrt{D}}}\sum_{i=0}^\infty \tau^i b_i(x,s)\hat{\gamma}^i(s).
\end{equation}
The functions $a_i(x,s)$ and $b_i(x,s)$ are determined applying the series $\sqrt{1+u}=1+u/2-u^2/8+u^3/16-\ldots$, $|u|<1$, and ${\rm e}^{-u}=\sum_{n=0}^\infty (-1)^n u^n/n!$. In the following, we use the approximation of $\hat{P}$ and $\hat{F}$ including four leading terms in the series.
For $0\leq i\leq 3$ we have
\begin{eqnarray}
a_0(x,s)=1,\;a_1(x,s)=\frac{1}{2}\Big(1-\frac{|x|s^{\alpha/2}}{\sqrt{D}}\Big),\label{eq40}\\
a_2(x,s)=\frac{1}{8}\Big(-1-\frac{|x|s^{\alpha/2}}{\sqrt{D}}+\frac{|x|^2 s^\alpha}{D}\Big),\label{eq41}\\
a_3(x,s)=\frac{1}{16}\Big(1+\frac{|x|s^{\alpha/2}}{\sqrt{D}}-\frac{|x|^{3/2} s^{3\alpha/2}}{3D^{3/2}}\Big),\label{eq42}
\end{eqnarray}
\begin{eqnarray}
b_0(x,s)=1,\;b_1(x,s)=-\frac{|x|s^{\alpha/2}}{2\sqrt{D}},\label{eq43}\\
b_2(x,s)=\frac{|x|s^{\alpha/2}}{8\sqrt{D}}\Big(1+\frac{|x|s^{\alpha/2}}{\sqrt{D}}\Big),\label{eq44}\\
b_3(x,s)=\frac{-|x|s^{\alpha/2}}{16\sqrt{D}}\Big(12+\frac{|x|s^{\alpha/2}}{\sqrt{D}}+\frac{|x|^2 s^\alpha}{3D}\Big).\label{eq45}
\end{eqnarray}

From Eqs. (\ref{eq20}) and (\ref{eq32}) we get
\begin{equation}\label{eq46}
\hat{\sigma}^2(s)=\frac{2D}{s^{1+\alpha}(1+\tau\hat{\gamma}(s))}.
\end{equation}
The series representation of the above function, with respect to $\tau$, is
\begin{equation}\label{eq47}
\hat{\sigma^2}(s)=\frac{2D}{s^{1+\alpha}}\sum_{i=0}^\infty \tau^i (-\hat{\gamma}(s))^i.
\end{equation}

\section{CTSE with the additional Caputo fractional time derivative\label{secV}}

Let us assume
\begin{equation}\label{eq48}
\hat{\gamma}(s)=s^\kappa,\; 0<\kappa<1.
\end{equation} 
In terms of the Laplace transform we get
\begin{equation}\label{eq49}
(\tau s^\kappa +1)\big[s\hat{P}(x,s)-P(x,0)\big]
=Ds^{1-\alpha}\frac{\partial^2 \hat{P}(x,s)}{\partial x^2}.
\end{equation}
Using the relation 
\begin{equation}\label{eq50}
\mathcal{L}\Bigg[\frac{^C\partial^\alpha f(t)}{\partial t^\alpha}\Bigg](s)=s^\alpha\hat{f}(s)-\sum_{i=0}^{n-1}s^{\alpha-1-i}f^{(i)}(0),
\end{equation}
where 
\begin{equation}\label{eq51}
\frac{^Cd^{\alpha} f(t)}{dt^\alpha}=\frac{1}{\Gamma(n-\alpha)}\int_0^t (t-t')^{n-\alpha-1}f^{(n)}(t')dt'
\end{equation}
is the Caputo fractional derivative (the natural number $n$ is defined as under Eq. (\ref{eq2})), and assuming $P^{(1)}(x,0)=0$, we get
\begin{equation}\label{eq52}
\tau\frac{^C\partial^{1+\kappa} P(x,t)}{\partial t^{1+\kappa}}+\frac{\partial P(x,t)}{\partial t}=D\frac{^{RL}\partial^{1-\alpha}}{\partial t^{1-\alpha}}\frac{\partial^2 P(x,t)}{\partial x^2}.
\end{equation}

To calculate the inverse Laplace transform we use the formula \cite{tkoszt2004}
\begin{eqnarray}\label{eq53}
\mathcal{L}^{-1}\left[s^\nu {\rm e}^{-as^\mu}\right](t)\equiv f_{\nu,\mu}(t;a)\\
=\frac{1}{t^{1+\nu}}\sum_{j=0}^\infty \frac{1}{j!\Gamma(-\nu-\mu j)}\left(-\frac{a}{t^\mu}\right)^j,\nonumber
\end{eqnarray}
$a,\mu>0$, $f_{\nu,\beta}$ is a special case of the Fox H--function (see Refs. \cite{mk,tkoszt2004} and the references cited therein),
\begin{eqnarray*}
  f_{\nu,\mu}(t;a)= \frac{1}{\mu a^{(1+\nu)/\mu}} H^{1 0}_{1 1}\left(\left.\frac{a^{1/\mu}}{t}\right|
    \begin{array}{cc}
      1 & 1 \\
      (1+\nu)/\mu & 1/\mu
    \end{array}
  \right)
  \;.
\end{eqnarray*}

From Eqs. (\ref{eq38})--(\ref{eq45}) and Eq. (\ref{eq53}) we get 
\begin{widetext}
\begin{eqnarray}\label{eq54}
P(x,t)=\frac{1}{2\sqrt{D}}\Bigg[\xi_{-1+\alpha/2}(|x|,t)
+\frac{\tau}{2}\Big(\xi_{-1+\kappa+\alpha/2}(|x|,t)-\frac{|x|}{\sqrt{D}}\xi_{-1+\kappa+\alpha}(|x|,t)\Big)\\
+\frac{\tau^2}{8}\Big(-\xi_{-1+2\kappa+\alpha/2}(|x|,t)
-\frac{|x|}{\sqrt{D}}\xi_{-1+2\kappa+\alpha}(|x|,t)+\frac{|x|^2}{D}\xi_{-1+2\kappa+3\alpha/2}(|x|,t)\Big)\nonumber\\
+\frac{\tau^3}{16}\Big(\xi_{-1+3\kappa+\alpha/2}(|x|,t)+\frac{|x|}{\sqrt{D}}\xi_{-1+3\kappa+\alpha}(|x|,t)-\frac{|x|^3}{3D^{3/2}}\xi_{-1+3\kappa+2\alpha}(|x|,t)\Big)\Bigg],\nonumber
\end{eqnarray}

\begin{eqnarray}\label{eq55}
F(t;x)=\xi_{0}(|x|,t)-\frac{\tau |x|}{2\sqrt{D}}\xi_{\kappa+\alpha/2}(|x|,t)+\frac{\tau^2 |x|}{8\sqrt{D}}\Big(\xi_{2\kappa+\alpha/2}(|x|,t)
+\frac{|x|}{\sqrt{D}}\xi_{2\kappa+\alpha}(|x|,t)\Big)\\
-\frac{\tau^3 |x|}{16\sqrt{D}}\Big(12\xi_{3\kappa+\alpha/2}(|x|,t)+\frac{|x|}{\sqrt{D}}\xi_{3\kappa+\alpha}(|x|,t)+\frac{|x|^2}{3D}\xi_{3\kappa+3\alpha/2}(|x|,t)\Big),\nonumber
\end{eqnarray}
\end{widetext}
where  
\begin{equation}\label{eq56}
\xi_\nu(x,t)\equiv f_{\nu,\alpha/2}\left(t;\frac{x}{\sqrt{D}}\right).
\end{equation} 
A special case of this function for $\nu=-1+\alpha/2$ is called the Mainardi function.

Assuming $\tau s^\kappa <1$ we obtain
\begin{equation}\label{eq57}
\hat{\sigma^2}(s)=\frac{2D}{s^{1+\alpha}}\sum_{i=0}^\infty (-\tau)^i s^{\kappa i}.
\end{equation}
The inverse Laplace transform method described in Appendix D provides
\begin{equation}\label{eq58}
\sigma^2(t)=2Dt^\alpha\sum_{i=0}^\infty \frac{(-\tau)^i}{\Gamma(1+\alpha-i\kappa)t^{i\kappa}}.
\end{equation}
In the long time limit the functions Eqs. (\ref{eq55}) and (\ref{eq58}) generate the following relations
\begin{equation}\label{eq59}
F_R(t\rightarrow\infty;x)=A_{\alpha,\kappa}\frac{\tau}{t^\kappa},
\end{equation}
\begin{equation}\label{eq60}
\sigma^2_R(t\rightarrow\infty)=B_{\alpha,\kappa}\frac{\tau}{t^\kappa},
\end{equation}
where $A_{\alpha,\kappa}=-\Gamma(-\alpha/2)/[2\Gamma(-\kappa-\alpha/2)]$ and $B_{\alpha,\kappa}=\Gamma(1+\alpha)/\Gamma(1+\alpha-\kappa)$.
Both above functions show that the Cattaneo effect disappears in the limit of long time as $\tau/t^\kappa$.

\begin{figure}[htb]
\centering{%
\includegraphics[scale=0.45]{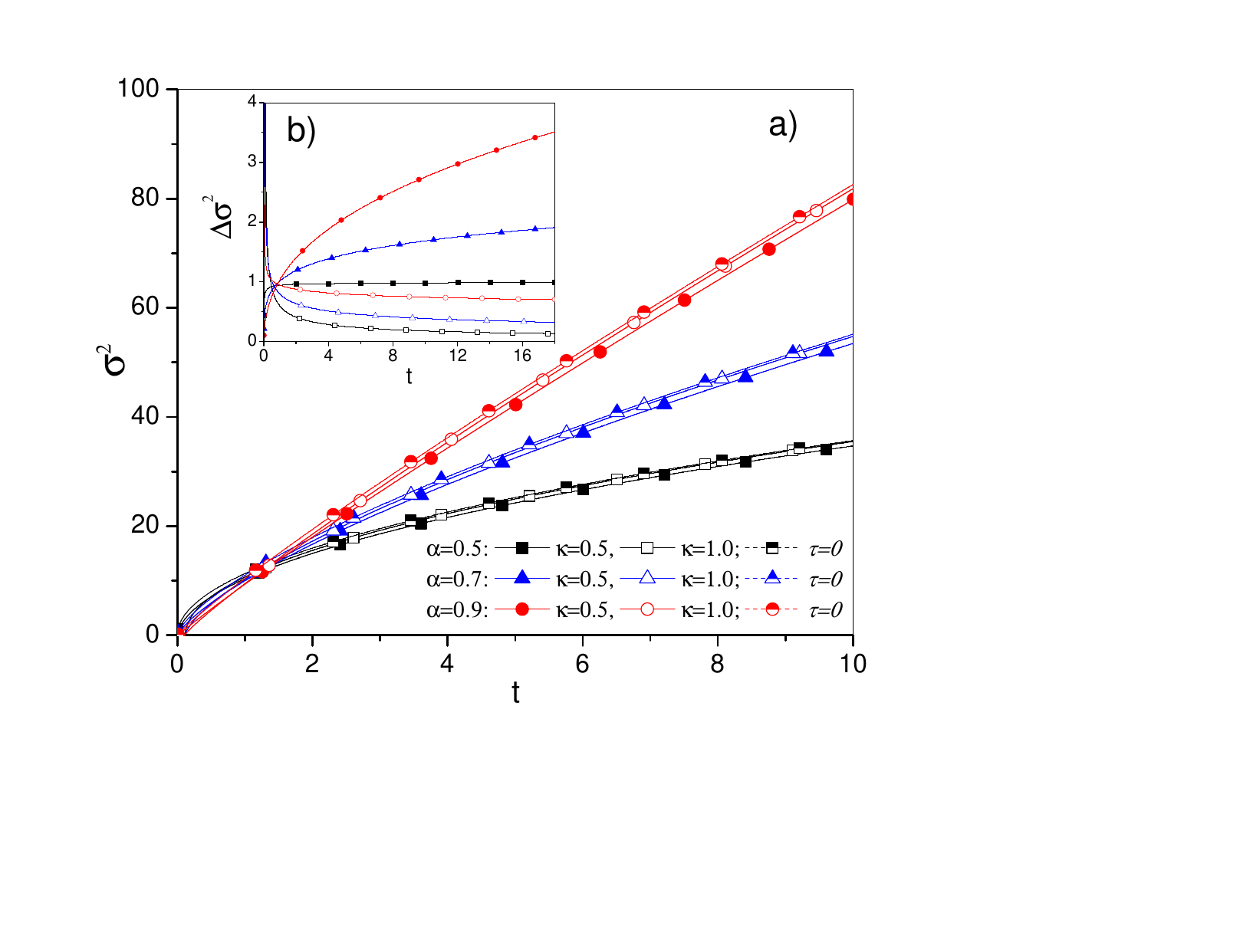}}
\caption{Plots of the time evolution of the MSD $\sigma^2$ Eq. (\ref{eq58}) (panel $a$) and the function $\Delta\sigma^2=\sigma_{\tau=0}^2-\sigma^2$ for the values of parameters in the legend (panel $b$), here $D=5$.}
\label{fig1}
\end{figure}

\begin{figure}[htb]
\centering{%
\includegraphics[scale=0.45]{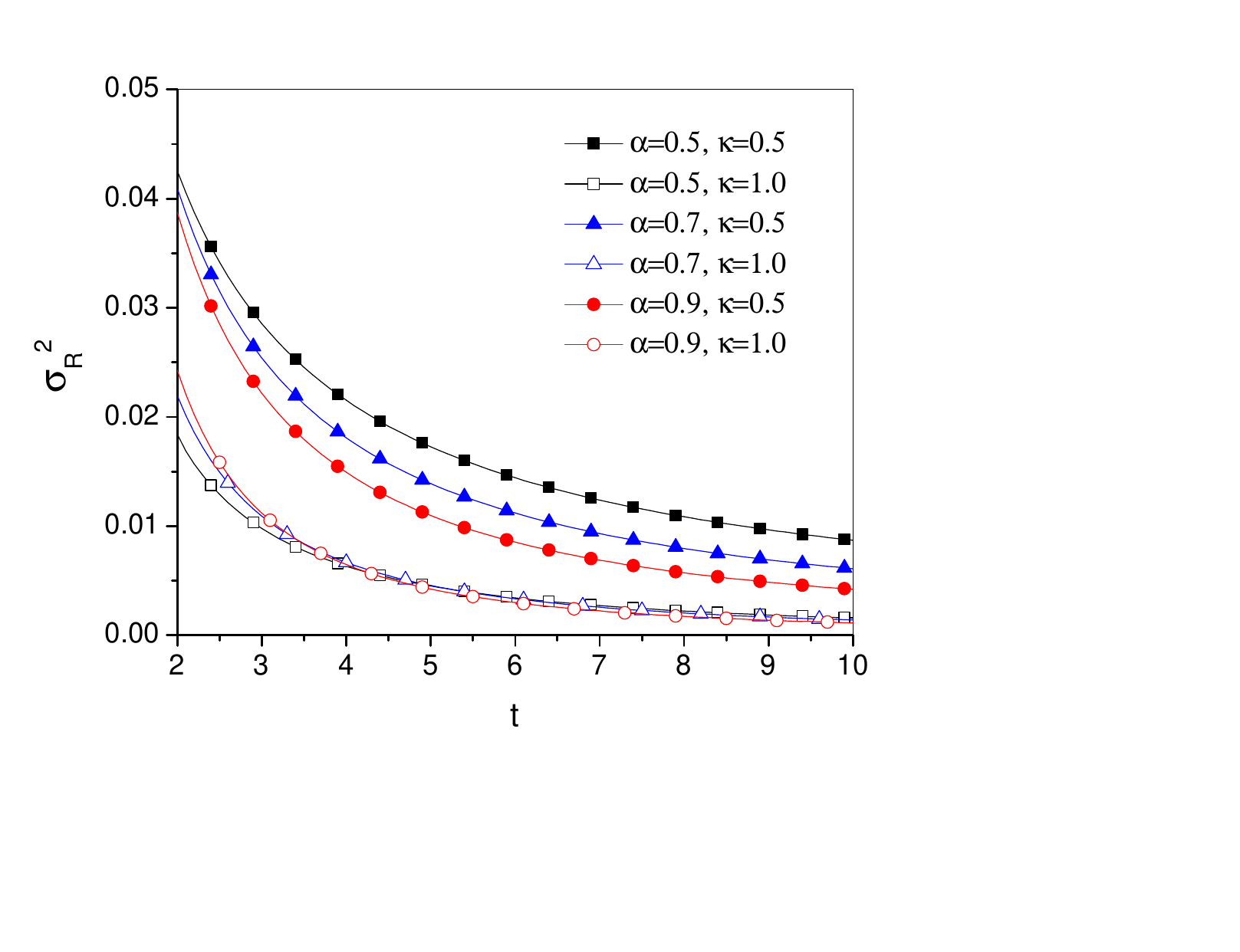}}
\caption{Time evolution of the relative MSD for the parameters given in the legend, $D=5$.}
\label{fig2}
\end{figure}

\begin{figure}[htb]
\centering{%
\includegraphics[scale=0.45]{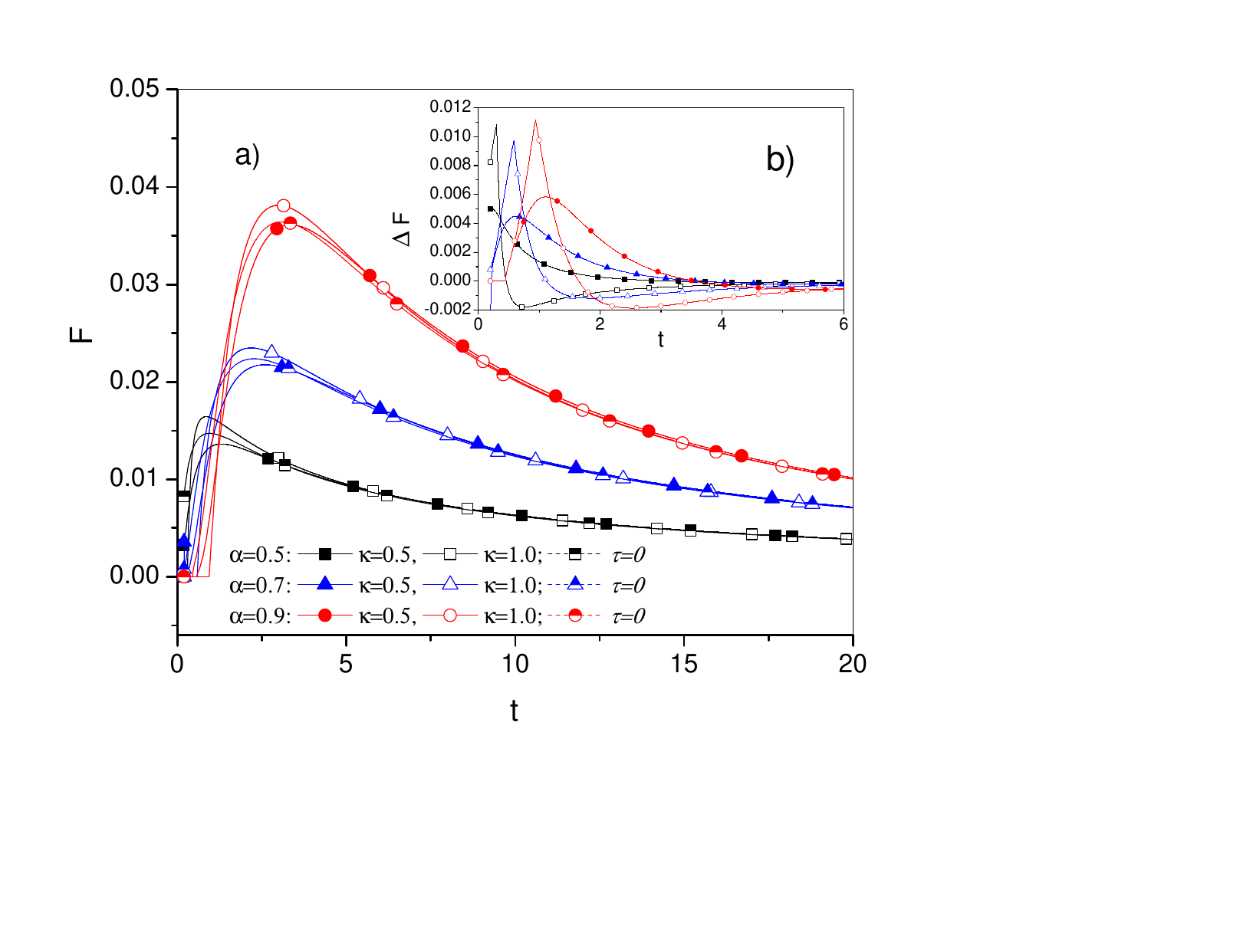}}
\caption{Time evolution of the first passage time distribution (panel $a$) and the function $\Delta F=F_{\tau=0}-F$ (panel $b$) for the parameters given in the legend, $x=10$ and $D=5$.}
\label{fig3}
\end{figure}

\begin{figure}[htb]
\centering{%
\includegraphics[scale=0.45]{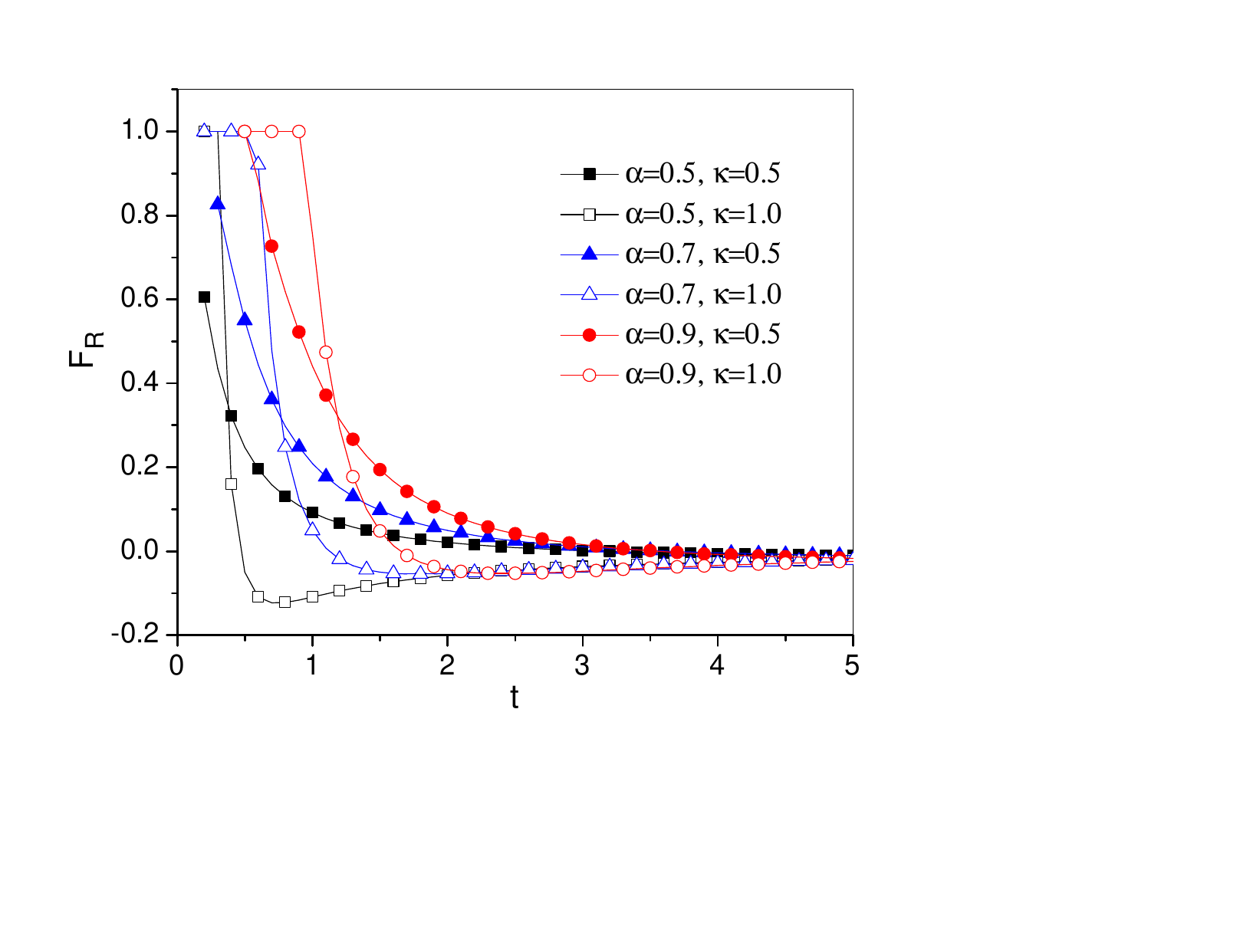}}
\caption{Time evolution of the relative first passage time for the parameters given in the legend, $x=10$ and $D=5$.}
\label{fig4}
\end{figure}

\begin{figure}[htb]
\centering{%
\includegraphics[scale=0.45]{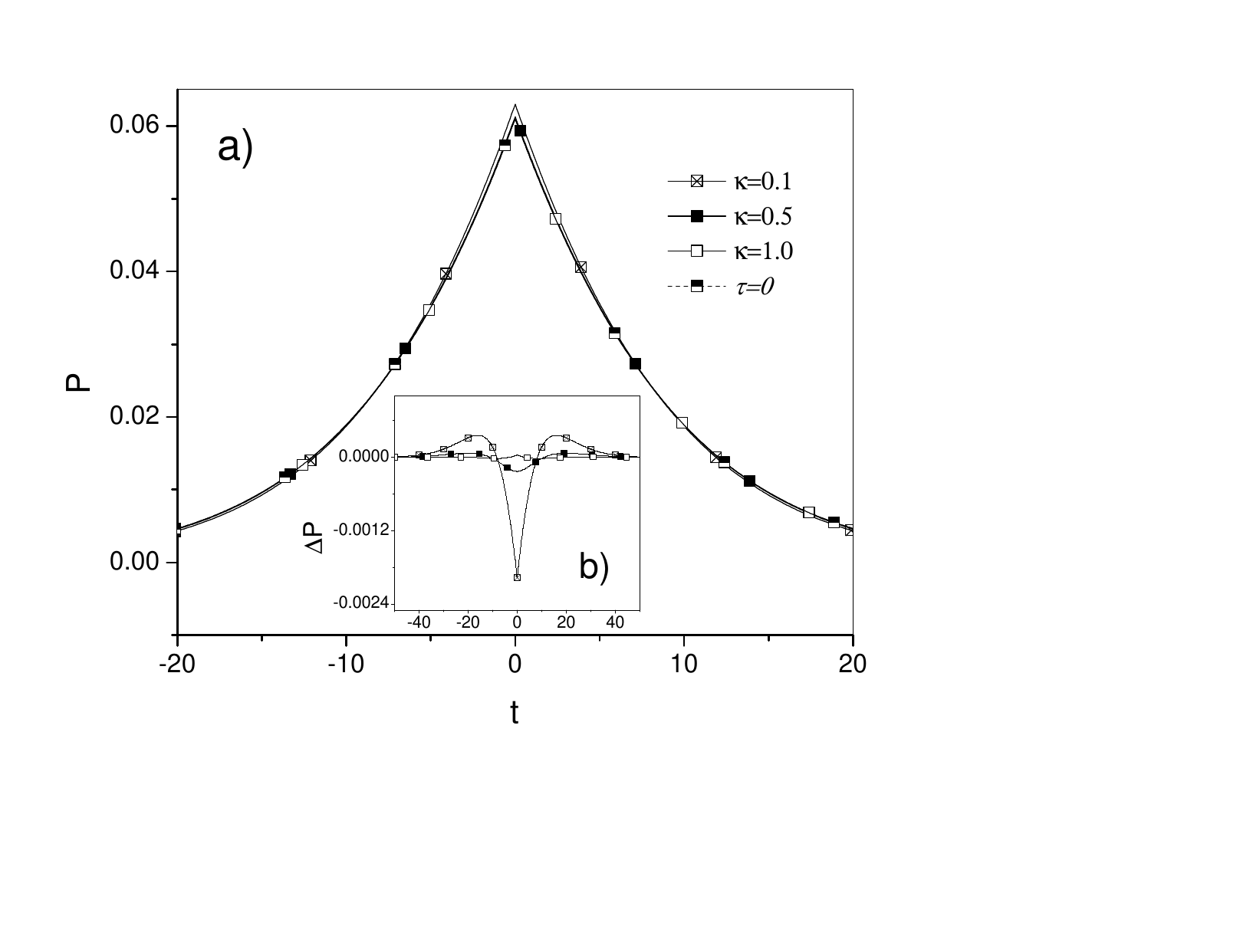}}
\caption{Plots of Green's functions for $\alpha=0.5$ (panel $a$) and functions $\Delta P=P_{\tau=0}-P$ for parameters given in the legend,
$t=20$ and $D=10$.}
\label{fig5}
\end{figure}

\begin{figure}[htb]
\centering{%
\includegraphics[scale=0.45]{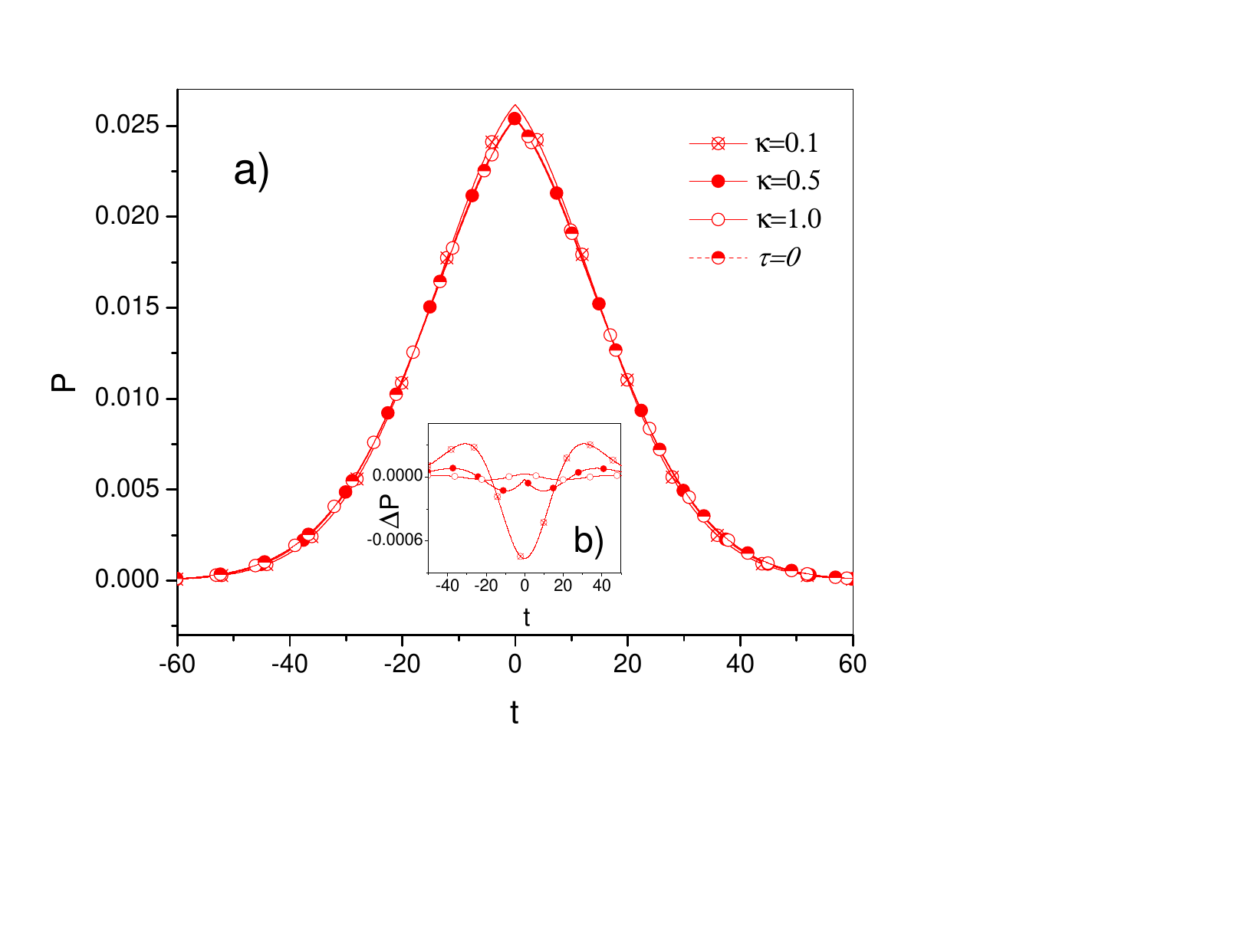}}
\caption{Plots of Green's function for $\alpha=0.9$, additional description is the same as for Fig. \ref{fig5}.}
\label{fig6}
\end{figure}

\begin{figure}[htb]
\centering{%
\includegraphics[scale=0.45]{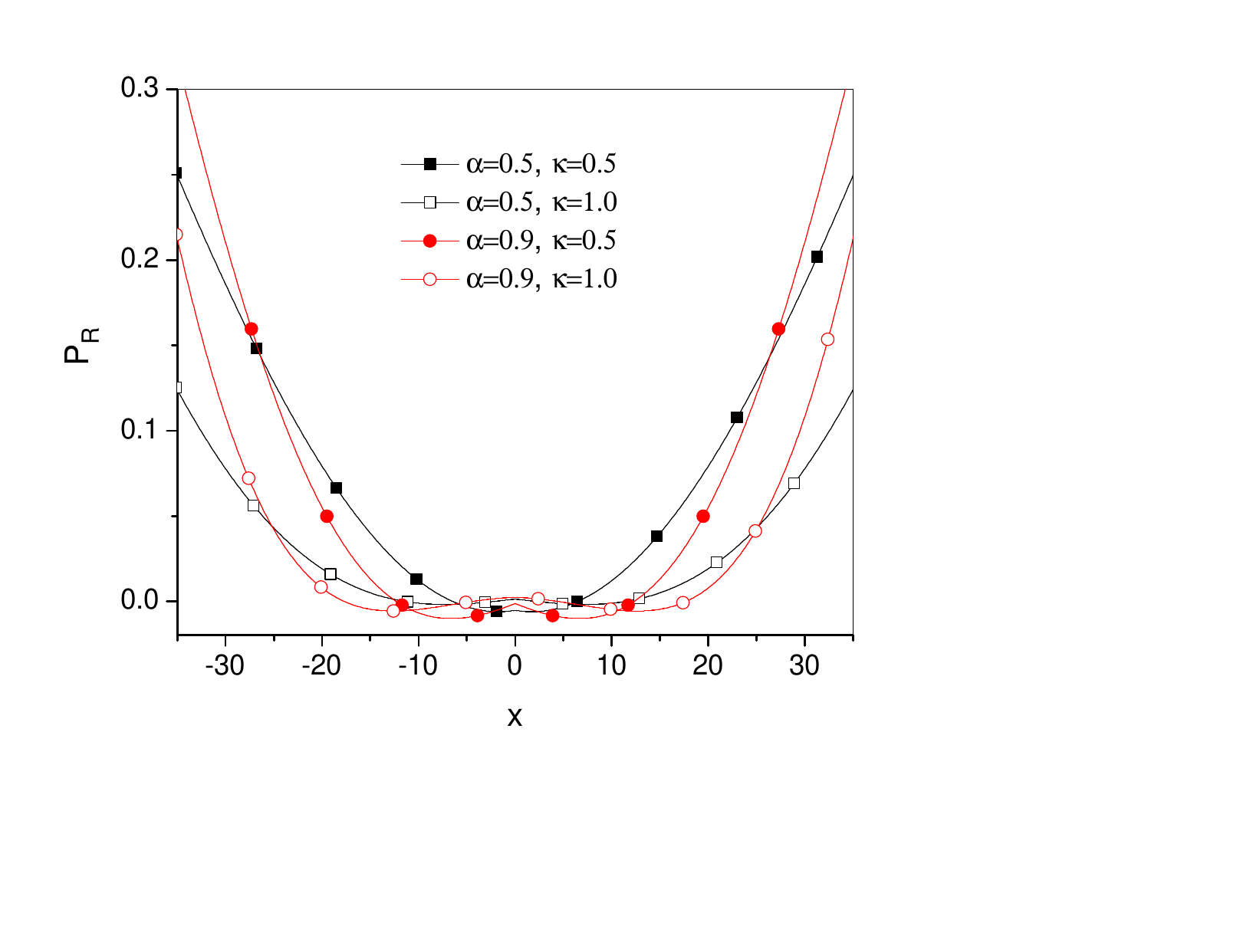}}
\caption{The relative Green's function $P_R$ for $\alpha$ and $\kappa$ given in the legend, $t=20$ and $D=10$.}
\label{fig7}
\end{figure}

The figures \ref{fig1}--\ref{fig7} show the plots of the functions describing subdiffusion with the Cattaneo effect. The functions are plotted for $\tau=0.1$, values of variables and parameters are given in arbitrarily chosen units. In Fig. \ref{fig1}, panel $a$, the time evolution of MSD for different parameters $\alpha$ and $\kappa$ are shown, the plots for $\tau=0$ show the function for the process without the Cattaneo effect. Panel $b$ shows the plots of the function $\Delta\sigma^2(t)=\sigma^2_{\tau=0}(t)-\sigma^2(t)$. The evolutions of the relative change of MSD $\sigma^2_R(t)$ are in Fig. \ref{fig2}. The distributions of the first passage time $F(t;x)$ (panel $a$) for the parameters as in Fig. \ref{fig1} and the differences $\Delta F(t;x)=F_{\tau=0}(t;x)-F(t;x)$ (panel $b$) are presented in Fig. \ref{fig3}. The relative changes of FPT, $F_R(t;x)$, are shown in Fig. \ref{fig4}. We note that, by Eqs. (\ref{eq27}) and (\ref{eq28}), the properties of the flux $J[P]$ are the same as the properties of the function $F$. The plots of the Green's function for different values of the parameter $\alpha$ are shown in Figs. \ref{fig5} and \ref{fig6} (panel $a$), and their differences in panel $b$. The relative values of the Green's functions are shown in Fig. \ref{fig7}. Figs. \ref{fig1}--\ref{fig6} suggest that the Cattaneo effect disappears over time, the main influence on the functions is the parameter $\kappa$. In Fig. \ref{fig7} we see that the Cattaneo effect increases with moving away from the initial position of a particle. This effect causes that finding the particle far from its initial position is much less likely when the process is described by the Cattaneo-type subdiffusion equation.

The Laplace transform of $R$ that generates the Cattaneo effect in Eq. (\ref{eq52}) is 
\begin{equation}\label{eq60a}
\hat{R}(s)=\frac{1}{1+\tau s^\kappa},
\end{equation} 
$\kappa\in(0,1)$. Due to the formula \cite{gorenflo}
\begin{equation}\label{eq60c}
\mathcal{L}\left[t^{\beta-1}E_{\rho,\beta}(\eta t^\rho)\right](s)=\frac{s^{\rho-\beta}}{s^\rho-\eta},
\end{equation}
$|\eta s^{-\rho}|<1$, where 
\begin{equation}\label{eq60d}
E_{\rho,\beta}(u)=\sum_{j=0}^\infty \frac{u^j}{\Gamma(\rho j+\beta)}
\end{equation}
is the two--parametric Mittag--Leffler function, we get
\begin{equation}\label{eq60e}
R(t)=\frac{1}{\tau t^{1-\kappa}}E_{\kappa,\kappa}\left(-\frac{t^\kappa}{\tau}\right).
\end{equation} 
However, Eq. (\ref{eq60c}) is valid for $s>1/\tau^{1/\rho}$, which causes the function $R(t)$ Eq. (\ref{eq60e}) to be defined for short time $t<t_1$, where $t_1$ is a certain parameter. For $s<1/\tau^{1/\rho}$ we have
\begin{equation}\label{eq60f}
\hat{R}(s)=\sum_{n=0}^\infty (-\tau)^n s^{n\kappa}.
\end{equation}
Using the method described in Appendix D we obtain 
\begin{equation}\label{eq60g}
R(t)=-\frac{\tau}{t^{1+\kappa}}\tilde{E}_{-\kappa,-\kappa}\left(-\frac{\tau}{t^\kappa}\right),
\end{equation}
$t>t_2$, where
\begin{equation}\label{eq60h}
\tilde{E}_{-\rho,-\beta}(u)=\sum_{j=0}^\infty \frac{u^j}{\Gamma(-\rho j-\beta)},
\end{equation}
$\rho,\beta>0$. The parameters $t_1$ and $t_2$ are determined from additional considerations, see Appendix D and the discussion in Ref. \cite{tk2023}.
The distribution $R(t)$ has a heavy tail,
\begin{equation}\label{eq60b} 
R(t\rightarrow\infty)\rightarrow \frac{-\tau}{\Gamma(-\kappa)t^{1+\kappa}},
\end{equation} 
the mean value of delay time is infinite. When $\kappa=1$, the average delay time is equal to $\tau<\infty$.
\begin{figure}[htb]
\centering{%
\includegraphics[scale=0.45]{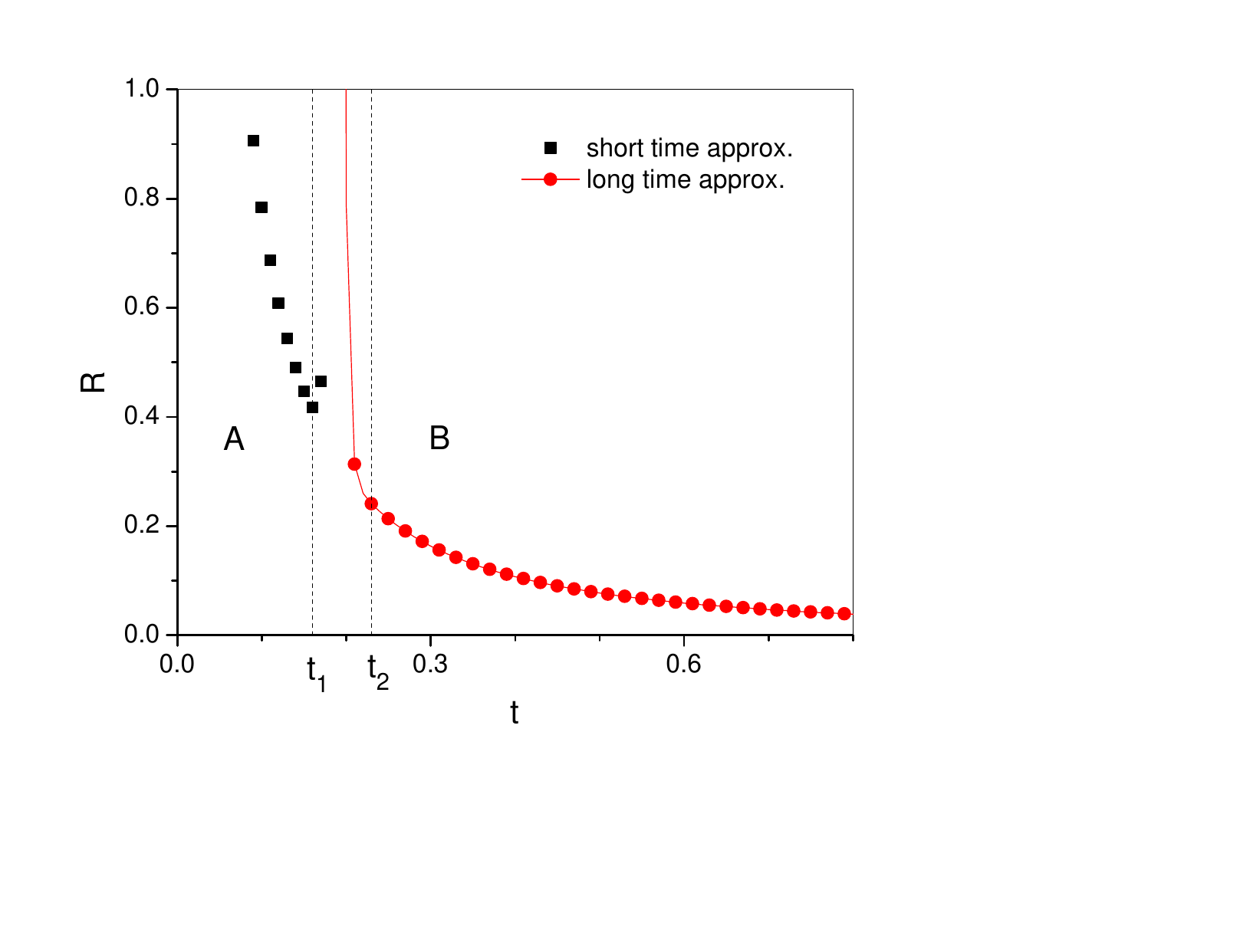}}
\caption{Plots of the function $R(t)$ for $\kappa=0.5$ and $\tau=0.1$ in the short time approximation Eq. (\ref{eq60e}) for $t\in(0,t_1)$ and in the long time approximation Eq. (\ref{eq60g}) for $t>t_2$, here $t_1=0.16$ and $t_2=0.23$.}
\label{fig1a}
\end{figure}
An example plot of the function $R(t)$ is shown in Fig. \ref{fig1a}. The dashed vertical lines show the time intervals for the short- and long-time approximations. The calculations include 20 leading terms of the series appearing in Eqs. (\ref{eq60e}) and (\ref{eq60g}).

Within the continuous time random walk (CTRW) model \cite{mk,ks}, for subdiffusion the probability density $\psi$ of the waiting time for a molecule to jump is heavy-tailed, $\psi(t\rightarrow\infty)\sim 1/t^{1+\alpha}$ with $\alpha\in(0,1)$, the mean value of this time is infinite. For normal diffusion there is $\psi(t)= {\rm e}^{-\lambda t}/\lambda$, the average waiting time for a molecule to jump equals $\lambda$. It is interesting to consider the different cases in which $\psi$ and $R$, independently of each other, have heavy tails or generate finite average values of $t$.

(a) Assuming that both distributions $\psi$ and $R$ have finite mean values we get Cattaneo normal diffusion equation Eq. (\ref{eq15}). In the long time limit we have
\begin{equation}\label{eq74a}
\sigma^2(t)\approx 2Dt\Big(1-\frac{\tau}{t}\Big).
\end{equation}

(b) Let $\psi$ be heavy-tailed and $R$ have a finite mean value. Then,
\begin{equation}\label{eq75}
\tau\frac{\partial^2 P(x,t)}{\partial t^2}+\frac{\partial P(x,t)}{\partial t}=D\frac{^{RL}\partial^{1-\alpha}}{\partial t^{1-\alpha}}\frac{\partial^2 P(x,t)}{\partial x^2},
\end{equation}
and for long time there is
\begin{equation}\label{eq76}
\sigma^2(t)\approx\frac{2Dt^\alpha}{\Gamma(1+\alpha)}\Big(1-\frac{\tau\alpha}{t}\Big).
\end{equation}

(c) Let $\psi$ have a finite mean value and $R$ be heavy-tailed. Then,
\begin{equation}\label{eq77}
\tau\frac{^C\partial^{1+\kappa} P(x,t)}{\partial t^{1+\kappa}}+\frac{\partial P(x,t)}{\partial t}=D\frac{\partial^2 P(x,t)}{\partial x^2}.
\end{equation}
In the limit of long time we get
\begin{equation}\label{eq78}
\sigma^2(t)\approx 2Dt\Big(1-\frac{\tau}{\Gamma(2-\kappa)t}\Big).
\end{equation}

(d) When both $\psi$ and $R$ are heavy-tailed, CTSE is expressed by Eq. (\ref{eq52}).
In the limit of long time we have
\begin{equation}\label{eq79}
\sigma^2(t)\approx\frac{2Dt^\alpha}{\Gamma(1+\alpha)}\Big(1-\frac{\tau\Gamma(1+\alpha)}{\Gamma(1+\alpha-\kappa)t^\kappa}\Big).
\end{equation}

Eq. (\ref{eq75}) is equivalent to Eq. GCE III, and Eq. (\ref{eq52}) coincides, for $\alpha=\kappa$, with Eq. GCE I in Ref. \cite{compte1997}. In the above mentioned paper, Cattaneo subdiffusion equations have been derived phenomenologically by combining the continuity equation with the constitutive flux equation, in both of which natural order time derivatives have been changed, in different ways, to fractional time derivatives. We note that $\sigma^2(t)$ approaches the normal diffusion relation $\sigma^2(t\rightarrow\infty)=2Dt$ when $\psi$ provides a finite value, independently of $R$. When $\psi$ is heavy-tailed, MSD approaches the ordinary subdiffusion function $\sigma^2(t\rightarrow\infty)=2Dt^\alpha/\Gamma(1+\alpha)$, this result is also independent of $R$. We conclude that the distribution $\psi$ has a stronger effect on the type of diffusion than $R$. 

\section{CTSE with an additional operator generated by a slowly varying function\label{secVI}}

A slowly varying function at infinity $\mathcal{\omega}(t)$ fulfils the relation
\begin{equation}\label{eq61}
\frac{\mathcal{\omega}(at)}{\mathcal{\omega}(t)}\rightarrow 1
\end{equation}
when $t\rightarrow\infty$ for any $a>0$. Functions that have finite limits when $t\rightarrow\infty$ and logarithmic functions are examples of slowly varying functions.

In the following, we use the strong Tauberian theorem to determine the inverse Laplace transform in the long-time limit \cite{hughes}:
{\it if $\phi(t)\geq 0$, $\phi(t)$ is ultimately monotonic like $t\rightarrow\infty$, $\mathcal{\omega}$ is slowly varying at infinity, and $0<\rho<\infty$, then each of the relations
\begin{equation}\label{eq62}
\hat{\phi}(s)\approx\frac{\mathcal{\omega}(1/s)}{s^\rho}
\end{equation}
as $s\rightarrow 0$ and
\begin{equation}\label{eq63}
\phi(t)\approx\frac{\mathcal{\omega}(t)}{\Gamma(\rho)t^{1-\rho}}
\end{equation}
as $t\rightarrow\infty$ implies the other.}

Let $\hat{\gamma}(s)\equiv\mathcal{\omega}(1/s)$ be a slowly varying function. When $s\rightarrow 0$, the Laplace transform of Green's function Eq. (\ref{eq36}) is approximated as 
\begin{eqnarray}\label{eq64}
\hat{P}(x,s)=\frac{\sqrt{1+\tau\mathcal{\omega}(1/s)}}{2\sqrt{D}s^{1-\alpha/2}}\\
\times\Bigg[1-\frac{s^{\alpha/2}|x|\sqrt{1+\tau\mathcal{\omega}(1/s)}}{\sqrt{D}}\Bigg].\nonumber
\end{eqnarray}
Using the strong Tauberian theorem we get in the long time limit
\begin{eqnarray}\label{eq65}
P(x,t)=\frac{\sqrt{1+\tau\mathcal{\omega}(t)}}{2\sqrt{D}\Gamma(1-\alpha/2)t^{\alpha/2}}\\
\times\Bigg[1-\frac{\Gamma(1-\alpha/2)\sqrt{1+\tau\mathcal{\omega}(t)}|x|}{\sqrt{D}\Gamma(1-\alpha)t^{\alpha/2}}\Bigg].\nonumber
\end{eqnarray}
Keeping the leading term in the above equation and putting it to Eq. (\ref{eq21}) we get
\begin{eqnarray}\label{eq66}
F(t;x)=\frac{|x|}{2\sqrt{D}\Gamma(1-\alpha/2)\sqrt{1+\tau\mathcal{\omega}(t)}t^{\alpha/2}}\\
\times\Bigg[\frac{\alpha(1+\tau\mathcal{\omega}(t))}{2t}-\tau\mathcal{\omega}^{(1)}(t)\Bigg].\nonumber
\end{eqnarray}
Eqs. (\ref{eq29}) and (\ref{eq66}) provide
\begin{eqnarray}\label{eq67}
F_R(t;x)=1-\sqrt{1+\tau\mathcal{\omega}(t)}\\
+\frac{2\tau t\mathcal{\omega}^{(1)}(t)}{\alpha\sqrt{1+\tau\mathcal{\omega}(t)}}.\nonumber
\end{eqnarray}
From Eqs. (\ref{eq20}), (\ref{eq30}), and the strong Tauberian theorem we obtain
\begin{equation}\label{eq68}
\sigma^2(t)=\frac{2Dt^\alpha}{\Gamma(1+\alpha)(1+\tau\mathcal{\omega}(t))},
\end{equation}
and
\begin{equation}\label{eq69}
\sigma_R^2(t)=\frac{\tau\mathcal{\omega}(t)}{1+\tau\mathcal{\omega}(t)}.
\end{equation}
Eqs. (\ref{eq65})--(\ref{eq69}) have been obtained in the long time limit.

As an example we consider the CTSE with AO generated by 
\begin{equation}\label{eq70}
\hat{\gamma}(s)\equiv\mathcal{\omega}(1/s)=\frac{1}{(1+s)\;{\rm log}(1+1/s)},
\end{equation}
$a>0$. The above function is a slowly varying function as well as a complete Bernstein function, see Ch. 15.4 in Ref. \cite{schilling}, and meet the condition $\hat{\gamma}(0^+)=0$.
Since
\begin{equation}\label{eq71}
\mathcal{\omega}(t)=\frac{t}{(1+t)\;{\rm log}(1+t)},
\end{equation}
in the limit of long time we get
\begin{equation}\label{eq72}
F_R(t\rightarrow\infty;x)=-\frac{\tau}{2\;{\rm log}t},
\end{equation}
\begin{equation}\label{eq73}
\sigma_R^2(t\rightarrow\infty)=\frac{\tau}{{\rm log}t}.
\end{equation}
The mean delay time can be calculated by means of the formula $\left\langle t\right\rangle=-\hat{R}^{(1)}(0)=\tau\hat{\gamma}^{(1)}(0)$.
The above equation and Eq. (\ref{eq70}) provide $\left\langle t\right\rangle=\infty$.

\section{CTSE derived from the standard CTRW model\label{secVII}}

We have derived the Cattaneo--type equation assuming that the flux is not generated at the same time as the concentration gradient. The time shift between the flux and the concentration gradient is controlled by the function $R$. However, there are other models that provide the Cattaneo equation. The equation has been derived within the continuous time persistent random walk (CTPRW) model, see Refs. \cite{gorska2020,masoliver,masoliver2016} and the references cited therein. Unlike in the ordinary continuous time random walk (CTRW) model \cite{montroll1965}, in CTPRW model the probability of choosing the direction of a molecule jump depends on the direction of its previous jump. In this section we show that, under certain conditions imposed on $R$, the Cattaneo--type equation can be derived by means of the ordinary CTRW model with a properly defined distribution of waiting time for a molecule to jump. 

The ordinary CTRW model provides the following equation given in terms of the Laplace transform, see Appendix C,
\begin{equation}\label{eqVII1}
s\hat{P}(x,s)-P(x,0)=\frac{\epsilon^2 s\hat{\psi}(s)}{2[1-\hat{\psi}(s)]}\frac{\partial^2 \hat{P}(x,s)}{\partial x^2},
\end{equation}
where $\epsilon$ is the mean value of a single jump length. Comparing Eqs. (\ref{eq10}) and (\ref{eqVII1}) we obtain
\begin{equation}\label{eqVII2}
\hat{\psi}(s)=\frac{1}{1+\frac{\lambda s^\alpha}{\hat{R}(s;\tau)}},
\end{equation}
where $\lambda=\epsilon^2/2D$. Guided by the considerations presented in Ref. \cite{gorska2020} we conclude that the function $\psi$ Eq. (\ref{eqVII2}) is non-negative when 
\begin{equation}\label{eqVII2a}
\frac{s^\alpha}{\hat{R}(s;\tau)}\in\mathcal{CBF}.
\end{equation} 
Since $\hat{R}(0;\tau)=1$, the normalization condition $\hat{\psi}(0)=1$ is met.

From Eqs. (\ref{eq60a}) and (\ref{eqVII2}) we get
\begin{equation}\label{eqV2}
\hat{\psi}(s)=\frac{1}{1+\lambda s^\alpha+\lambda\tau s^{\alpha+\kappa}}.
\end{equation}
Due to properties (4), (10), and (11) in Appendix A $\hat{\psi}(s)\in\mathcal{CMF}$ when
\begin{equation}\label{eqVII2c}
\alpha+\kappa\leq 1.
\end{equation}
In further considerations we assume that the condition Eq. (\ref{eqVII2c}) is met.
The inversion Laplace transform of $\hat{\psi}(s)$ is calculated separately in different time regimes. Below, we present the inverse transforms for the short time corresponding to the condition $\lambda s^\alpha>1$ and $\tau s^\kappa>1$ and for the long time corresponding to conditions $\tau s^\kappa<1$ and $\lambda s^\alpha<1$. We would like to add that a more detailed analysis of this issue will be presented elsewhere \cite{kdfut}.

For $\lambda s^\alpha>1$ and $\tau s^\kappa>1$ we get
\begin{eqnarray}\label{eqV2a}
\hat{\psi}(s)=-\sum_{n=1}^\infty\frac{(-1)^n}{(\lambda\tau)^n s^{(\alpha+\kappa)n}}\sum_{m=0}^\infty {n+m-1 \choose m}\\
\times \frac{(-1)^m}{\tau^m s^{\kappa n}}, \nonumber
\end{eqnarray}
and for $\lambda s^\alpha<1$ and $\tau s^\kappa<1$, such that $\lambda s^\alpha(1+\tau s^\kappa)<1$, we obtain
\begin{equation}\label{eqV2b}
\hat{\psi}(s)=\sum_{n=0}^\infty (-1)^n\lambda^n s^{\alpha n}\sum_{m=0}^n{n \choose m}\tau^m s^{\kappa m}.
\end{equation}
Using the method described in Appendix D we get for short time, $t<t_{1,\tau\neq 0}$,
\begin{eqnarray}\label{eqV2c}
\psi(t)=-\frac{1}{t}\sum_{n=1}^\infty \left(-\frac{t^{\alpha+\kappa}}{\lambda\tau}\right)^n\sum_{m=0}^\infty \left(-\frac{1}{\tau}\right)^m\\
 \times {n+m-1 \choose m} \frac{t^{\kappa m}}{\Gamma(\alpha n+\kappa(n+m))}, \nonumber
\end{eqnarray}
and for long time $t>t_{2,\tau\neq 0}$,
\begin{eqnarray}\label{eqV2d}
\psi(t)=\frac{1}{t}\sum_{n=1}^\infty\left(-\frac{\lambda}{t^\alpha}\right)^n\sum_{m=0}^n {n \choose m}\\
\times\left(\frac{\tau}{t^\kappa}\right)^m\frac{1}{\Gamma(-\alpha n-\kappa m)}, \nonumber
\end{eqnarray}
where the parameters $t_{1,\tau\neq 0}$ and $t_{2,\tau\neq 0}$ are determined from additional considerations.

\begin{figure}[htb]
\centering{%
\includegraphics[scale=0.45]{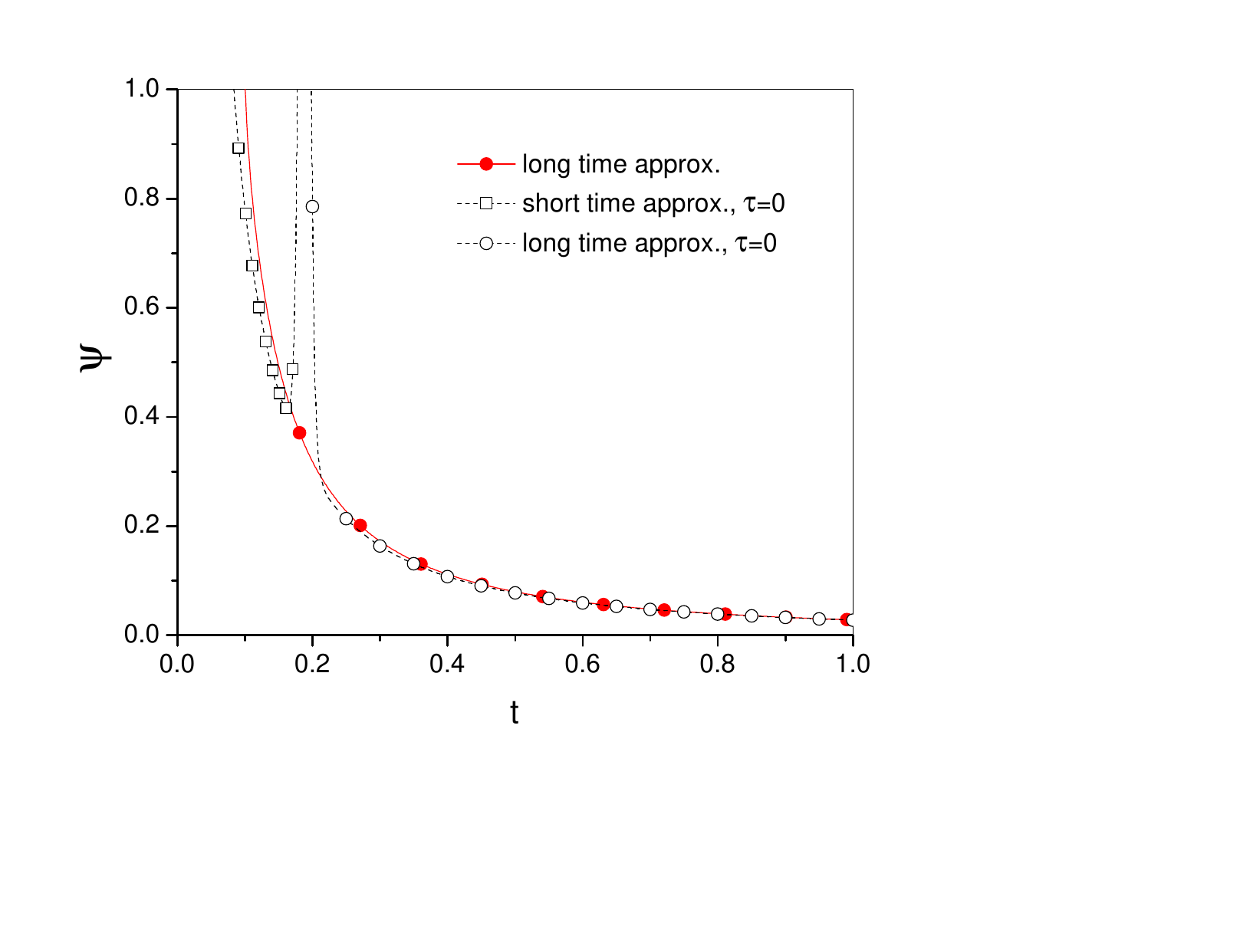}}
\caption{Plots of the function $\psi(t)$ for $\alpha=\kappa=0.5$ and $\lambda=\tau=0.1$ in the long time approximation Eq. (\ref{eqV2d}) (solid line) and the function $\psi_{\tau=0}(t)$ in the short time approximation Eq. (\ref{eqV2e}) for $t<0.16$ and in the long time approximation Eq. (\ref{eqV2f}) for $t>0.23$ (dashed lines).}
\label{fig1b}
\end{figure}

For a process without the Cattaneo effect, when $\tau=0$, $\hat{\psi}_{\tau=0}(s)$ takes a form similar to $\hat{R}(s)$ Eq. (\ref{eq60a}). Using the same method as in determining the function $R(t)$, we obtain
\begin{equation}\label{eqV2e}
\psi_{\tau=0}(t)=\frac{1}{\lambda t^{1-\alpha}}E_{\alpha,\alpha}\left(-\frac{t^\alpha}{\lambda}\right).
\end{equation}
for $t<t_{1,\tau=0}$ and 
\begin{equation}\label{eqV2f}
\psi_{\tau=0}(t)=-\frac{\lambda}{t^{1+\alpha}}\tilde{E}_{-\alpha,-\alpha}\left(-\frac{\lambda}{t^\alpha}\right),
\end{equation}
for $t>t_{2,\tau=0}$.
The parameters $t_{1,\tau=0}$ and $t_{2,\tau=0}$ can be different from $t_{1,\tau\neq 0}$ and $t_{2,\tau\neq 0}$.

The plot of the function $\psi(t)$ for $\alpha=\kappa=0.5$ and $\lambda=\tau=0.1$ is shown in Fig. \ref{fig1b}. The time at which the short-time approximation is valid is estimated as $t^{\alpha+\kappa}/(\lambda\tau)<1$ (see Eq. (\ref{eqV2c})), which gives $t<0.01$. The plot for this approximation is not shown in Fig. \ref{fig1b}. The long-time approximation works for $t>0.01$. The plot of $\psi(t)$ is compared with the plots of $\psi_{\tau=0}(t)$ obtained within the short-time and long-time approximations. The function $\psi(t)$ lags $\psi_{\tau=0}(t)$, this lag decreasing with time.

\section{Final remarks\label{secVIII}}

The Cattaneo-type subdiffusion equation (CTSE) describes subdiffusion with involved Cattaneo effect which consists in delaying the activation of the ordinary subdiffusion flux by a random delay time. This effect is controlled by the delay time probability density $R$. The main remarks are as follows.
 
(1) Different anomalous diffusion processes have been described by differential-integral equations with memory kernel (MK). The examples are equations generated by distributed order MK \cite{sandev2015,sandev2015a,sandev2017,sandev2018,sandev2019}, power--law MK \cite{sandev2015}, and truncated power law MK \cite{sandev2017}, see also \cite{metzler1998,sandev2015a,sandev2018,sandev2019,chechkin2021}. The general form of CTSE, Eq. (\ref{eq11}), is a special case of an equation with MK.   
In our model MK Eq. (\ref{eq12}) is controlled by the probability density of delay time $R$. Then, MK is limited by the conditions 1-3 in Sec. \ref{secII} imposed on the function $R$.
When $\hat{R}(s;\tau)=1/[1+\tau\hat{\gamma}(s)]$, we get the Cattaneo--type subdiffusion equation that differs from the ordinary subdiffusion one by an integro--differential additional operator (AO) with kernel $\gamma(t)$ Eq. (\ref{eq34}). When $\hat{\gamma}(s)=s^\kappa$ with $0<\kappa<1$, the AO is the Caputo fractional time derivative of the order $1+\kappa$.

(2) The functions $\sigma^2_R$ and $F_R$ show the influence of Cattaneo effect on subdiffusion.
For the process described by the CTSE with additional operator being the Caputo fractional time derivative there is $\sigma^2_R\sim F_R\sim 1/t^\kappa$, $0<\kappa<1$, while for the equation with additional operator generated by a slowly varying function $\hat{\gamma}$ Eq. (\ref{eq70}) we have $F_R\sim\sigma^2_R\sim 1/{\rm log}t$. The examples suggest that the Cattaneo effect controlled by AO with a slowly varying function decays much slower than when AO is the Caputo fractional derivative.

(3) The subdiffusion process is generated by a heavy-tailed distribution of the waiting time for a molecule to jump $\psi_{\tau=0}$, which is controlled by the parameter $\alpha$. In our considerations, the distribution of $R$ generating the Cattaneo effect is independent of $\psi_{\tau=0}$, and it can also have a heavy tail. As is has been shown in Sec. \ref{secV}, for long times the subdiffusion effect is dominant over the Cattaneo effect.

(4) We have not considered functions describing subdiffusion with the Cattaneo effect in a short time limit. This is due to the lack of an unambiguous interpretation of this effect when $t\rightarrow 0$, which has been justified in Sec. \ref{secI}. In the model considered in Sec. \ref{secV}, in the short time limit we get $\sigma^2(t)=2Dt^{\alpha+\kappa}/[\tau\Gamma(1+\alpha+\kappa)]$. When $\alpha+\kappa>1$ this relation suggests superdiffusion. The main problem is how to justify the occurrence of superdiffusion in a subdiffusive medium. It is difficult to assume that, for example, large molecules would move superdiffusively in a dense gel in the initial short time period, when their movement is very hindered and subdiffusion has been experimentally observed there, see Ref. \cite{kdm}. This remark confirms the ambiguity of the physical interpretation of the process described by the CTSE in the short-time limit.

(5) An example of a process in which the Cattaneo effect may be important is subdiffusion (or normal diffusion when $\alpha=1$) of antibiotics in a bacterial biofilm. The defense mechanisms of bacteria against the action of antibiotics can change the random walk of antibiotic molecules \cite{aot,mot}. It may also be interesting to use the Cattaneo-type equation in modelling the spread of an epidemic. The first arrival of an infected object to some point may create new source of infection. Since the Cattaneo effect changes the distribution of $F$, the use of Eqs. (\ref{eq11}) or (\ref{eq34}) (with possibly added reaction terms in both equations) may give qualitatively different results when compared to the ones derived from ordinary subdiffusion equation.

\section*{Appendix A: Completely monotone and complete Bernstein functions}

A function $f:(0,\infty)\rightarrow \Re$ is a completely monotone function ($\mathcal{CMF}$) if $f$ is of class $C^\infty$ and $(-1)^nf^{(n)}(u)\geq 0$ for $n=0,1,2,\ldots$, where $f^{(n)}(u)=d^n f(u)/du^n$. The function $f$ is a complete Bernstein function ($\mathcal{CBF}$) if $f(u)/u$ is the Laplace transform of $\mathcal{CMF}$ restricted to the positive semiaxis. The Brernstein theorem provides that $\hat{F}(s)\in\mathcal{CMF}$ if and only if $f(t)\geq 0$ for $t\geq 0$, where $\hat{F}(s)=\int_0^\infty {\rm exp}(-st)f(t)dt$. Some properties of the functions are \cite{gorska2020,schilling}:
\begin{enumerate}
	\item	$f,g\in\mathcal{CMF},\;a,b\geq 0\Rightarrow af+bg\in\mathcal{CMF}$, \label{a1}
	\item $f,g\in\mathcal{CMF}\Rightarrow f\cdot g\in\mathcal{CMF}$, \label{a2}
	\item $f\in\mathcal{CMF}\;and\;g\in\mathcal{CBF}\Rightarrow f\circ g\equiv f(g)\in\mathcal{CMF}$, \label{a3}
	\item	$f,g\in\mathcal{CBF},\;a,b\geq 0\Rightarrow af+bg\in\mathcal{CBF}$,
	\item $f,g\in\mathcal{CBF}\Rightarrow (f^\alpha +g^\alpha)^{1/\alpha}\in\mathcal{CBF},\;\alpha\in[-1,1]/\{0\}$,
	\item $f,g\in\mathcal{CBF}\Rightarrow f^\alpha\cdot g^\beta\in\mathcal{CBF},\;\alpha,\beta\in(0,1),\;\alpha+\beta\leq 1$,
	\item $f(u)\in\mathcal{CBF} (f(u)\neq 0)\Leftrightarrow u/f(u)\in\mathcal{CBF}$,
	\item $f(u)={\rm exp}(-au)\in\mathcal{CMF}$, $a>0$,
	\item $f(u)=u^\mu\in\mathcal{CMF}\;when\;\mu\leq 0$,
	\item $f(u)=u^\mu\in\mathcal{CBF}\;when\;\mu\in[0,1]$,
	\item	$1/f\in\mathcal{CBF}\Rightarrow f\in\mathcal{CMF}$,
	\item $f(u)=1/[(1+u){\rm log}(1+1/u)]\in\mathcal{CBF}$.
\end{enumerate}

\section*{Appendix B: Derivation of Eq. (\ref{eq21})}

The time the molecule arrives at point $x$ for the first time is the same as the time of absorption of the molecule by the absorbing wall placed at that point. The Green's function for the system with the absorbing wall is
\begin{equation}\label{b1}
P_{abs}(x',t)=P(x',t)-P(2x-x',t),
\end{equation}
where $x'<x$, the initial molecule position is $x_0=0$, and $P$ is the Green's function for the unbounded homogeneous system without the wall. Since $P(x',t)\equiv P(-x',t)$, the probability $\mathcal{P}$ that a molecule has not been absorbed by time $t$ is
\begin{equation}\label{b2}
\mathcal{P}(t)=\int_{-\infty}^x P_{abs}(x',t)dx'=\int_{-x}^x P(x',t)dx'.
\end{equation}
The probability that a molecule will be absorbed in the time interval $(t,t+\Delta t)$ is
\begin{equation}\label{b3}
F(t;x)\Delta t=\mathcal{P}(t)-\mathcal{P}(t+\Delta t).
\end{equation}
Eqs. (\ref{b1})--(\ref{b3}) provide Eq. (\ref{eq21}) when $\Delta t\rightarrow 0$.

\section*{Appendix C: Derivation of Eq. (\ref{eqVII1})}

Let us suppose the probability density of the jump length $\Delta x$ reads 
\begin{equation}\label{c1}
\xi(\Delta x)=[\delta(\Delta x-\epsilon)+\delta(\Delta x+\epsilon)]/2, 
\end{equation}
$\epsilon<\infty$.
In terms of the Fourier transform $\tilde{f}(k)\equiv\mathcal{F}
[f(u)](k)=\int_{-\infty}^\infty {\rm exp}(iku)f(u)du$ we have $\tilde{\xi}(k)={\rm cos}(k\epsilon)$; when $k\epsilon\rightarrow 0$, there is
\begin{equation}\label{c2}
\tilde{\xi}(k)\approx 1-\frac{k^2\epsilon^2}{2}.
\end{equation}
Combining Eq. (\ref{c2}) with the standard equation obtained within the ordinary CTRW model \cite{montroll1965}
\begin{equation}\label{c3}
\hat{\tilde{P}}(k,s)=\frac{1-\hat{\psi}(s)}{s}\frac{1}{1-\tilde{\xi}(k)\hat{\psi}(s)},
\end{equation}
we get
\begin{equation}\label{c4}
s\hat{\tilde{P}}(k,s)-1=-\frac{\epsilon^2 sk^2}{2[1-\hat{\psi}(s)]}\hat{\tilde{P}}(k,s).
\end{equation}
Using the formulas $\mathcal{F}^{-1}[1](x)=\delta(x)\equiv P(x,0)$ and $\mathcal{F}^{-1}[-k^2\hat{\tilde{P}}(k,s)](x)=\partial^2 \hat{P}(x,s)/\partial x^2$ we get Eq. (\ref{eqVII1}).

Another way to derive Eq. (\ref{eqVII1}) is the particle random walk model based on the difference equation
\begin{equation}\label{c5}
P_{n+1}(x)=\frac{1}{2}P_n(x-\epsilon)+\frac{1}{2}P_n(x+\epsilon),
\end{equation}
where $P_n(x)$ is the probability density of finding the molecule at $x$ after making $n$ jumps. Combining Eq. (\ref{c5}) and the formulas $P(x,t)=\sum_{n=0}^\infty Q_n(t)P_n(x)$, where $Q_n(t)$ is the probability that a particle makes $n$ jumps in the time interval $(0,t)$ (in terms of the Laplace transform there is $\hat{Q}(s)=\hat{\psi}^n(s)(1-\hat{\psi}(s))/s$) \cite{montroll1965}, and $\partial^2 P(x,t)/\partial x^2\approx [P(x+\epsilon,t)+P(x-\epsilon,t)-2P(x,t)]/\epsilon^2$, we obtain Eq. (\ref{eqVII1}) in the limit of small $\epsilon$.

\section*{Appendix D: Calculating the inverse Laplace transform}

We describe the method used in this paper to determine the inverse Laplace transforms of certain functions. We do not directly determine the inverse transform $\mathcal{L}^{-1}[\hat{g}(s)](t)$, but the transform $\mathcal{L}^{-1}[{\rm e}^{-as^\mu}\hat{g}(s)](t)$, $a,\mu>0$, for which we apply Eq. (\ref{eq53}) and then calculate the limit $a\rightarrow 0^+$; the result is independent of the parameter $\mu$. For example, suppose $\hat{g}(s)$ has two different power series expansions for large and small values of the positive real parameter $s$,
\begin{equation}\label{d1}
\hat{g}(s)=\frac{1}{s^{c_1}}\sum_{n=0}^\infty\frac{a_n}{s^{c_2 n}},\;s>s_1,
\end{equation}
\begin{equation}\label{d2}
\hat{g}(s)=s^{d_1}\sum_{n=0}^\infty b_n s^{d_2 n},\;s<s_2,
\end{equation}
where $c_2,d_2>0$ and $s_1$, $s_2$ are positive numbers.
Let us consider the following series
\begin{equation}\label{d3}
{\rm e}^{-as^\mu}\hat{g}(s)=\frac{{\rm e}^{-as^\mu}}{s^{c_1}}\sum_{n=0}^\infty\frac{a_n}{s^{c_2 n}},\;s>s_1,
\end{equation}
\begin{equation}\label{d4}
{\rm e}^{-as^\mu}\hat{g}(s)={\rm e}^{-as^\mu} s^{d_1}\sum_{n=0}^\infty b_n s^{d_2 n},\;s<s_2,
\end{equation}
$a,\mu>0$.
Eqs. (\ref{eq53}) and (\ref{d3}) give
\begin{equation}\label{d5}
\mathcal{L}^{-1}[{\rm e}^{-as^\mu}\hat{g}(s)](t)=\sum_{n=0}^\infty a_n f_{-c_1-c_2n,\mu}(a;t),\;t<t_1,
\end{equation}
and Eqs. (\ref{eq53}) and (\ref{d4}) provide
\begin{equation}\label{d6}
\mathcal{L}^{-1}[{\rm e}^{-as^\mu}\hat{g}(s)](t)=\sum_{n=0}^\infty b_n f_{d_1+d_2n,\mu}(a;t),\;t>t_2.
\end{equation}
In the limit $a\rightarrow 0^+$ we have $f_{\nu,\mu}(t;a)\rightarrow 1/[\Gamma(-\nu)t^{1+\nu}]$; when $\nu=n$ is a natural number, $f_{n,\mu}(t;a)\rightarrow 0$. When $a\rightarrow 0^+$, Eqs. (\ref{d5}) and (\ref{d6}) are, respectively, 
\begin{equation}\label{d7}
g(t)=\sum_{n=0}^\infty a_n \frac{t^{c_1+c_2 n-1}}{\Gamma(c_1+c_2 n)},\;t<t_1,
\end{equation}
\begin{equation}\label{d8}
g(t)=\sum_{n=0}^\infty b_n \frac{1}{\Gamma(-d_1-d_2 n)t^{d_1+d_2 n+1}},\;t>t_2.
\end{equation}
The inequalities $s>s_1$ and $s<s_2$ do not uniquely determine the corresponding parameters $t_1$ and $t_2$.
Determining these parameters requires additional considerations. One such method is to determine the time intervals $I_1=(0,t_1)$ and $I_2=(t_2,\infty)$ in which the series converge and the functions defined in the second of these intervals can be interpreted as an extension of the function defined in the first interval.


\begin{thebibliography}{33}

\bibitem{lee2021} D. S. W. Lee, N. S. Wingreen, and C. P. Brangwynne, Chromatin mechanics dictates subdiffusion and coarsening dynamics of embedded condensates, Nat. Phys. {\bf 17}, 531 (2021).
\bibitem{bijeljic2011} B. Bijeljic, P. Mostaghimi, and M. J. Blunt, Signature of non--Fickian solute transport in complex heterogeneous porous media, Phys. Rev. Lett. {\bf 107}, 204502 (2011).
\bibitem{barkai2012} E. Barkai, Y. Garini, and R. Metlzer, Strange kinetics of single molecules in living cells, Phys. Today \textbf{65}, 29 (2012).
\bibitem{kdm} T. Koszto{\l}owicz, K. Dworecki, and S. Mr\'owczy\'nski, How to measure subdiffusion parameters, Phys. Rev. Lett. {\bf 94}, 170602 (2005).
\bibitem{km} T. Koszto{\l}owicz and R. Metzler, Diffusion of antibiotics through a biofilm in the presence of diffusion and absorption barriers, Phys. Rev. E {\bf 102}, 032408 (2020).
\bibitem{wyss1986} W. Wyss, The fractional diffusion equation, J. Math. Phys. {\bf 27}, 2782 (1986).
\bibitem{hilferanton} R. Hilfer and L. Anton, Fractional master equations and fractal time random walks, Phys. Rev. E {\bf 51}, R848 (1995).
\bibitem{compte} A. Compte, Stochastic foundations of fractional dynamics, Phys. Rev. E {\bf 53}, 4191 (1996).
\bibitem{mks} R. Metzler, J. Klafter, and I. M. Sokolov, Anomalous transport in external fields: Continuous time random walks and fractional diffusion equations extended, Phys. Rev. E {\bf 58}, 1621 (1998).
\bibitem{mk} R. Metzler and J. Klafter, The random walk's guide to anomalous diffusion: a fractional dynamics approach, Phys. Rep.  {\bf 339}, 1 (2000).
\bibitem{barkai2000} E. Barkai, R. Metzler, and J. Klafter, From continuous time random walks to the fractional Fokker-Planck equation, Phys. Rev. E {\bf 61}, 132 (2000).
\bibitem{skb} I. M. Sokolov, J. Klafter, and A. Blumen, Fractional kinetics, Phys. Today {\bf 55}, 11, 48 (2002).
\bibitem{klages2008} R. Klages, G. Radons, and I. M. Sokolov, {\it Anomalous Transport: Foundations and Applications} (Wiley, New York, 2008).
\bibitem{ks} J. Klafter and I. M. Sokolov, {\it First Step in Random Walks. From Tools to Applications} (Oxford UP, New York, 2011).
\bibitem{cattaneo} C. Cattaneo,  Sulla conduzione del calore. Attidel Seminario Matematico e Fisicodella Università di Modena \textbf{3}, 83 (1948).
\bibitem{zhmakin} A. I. Zhmakin, The zoo of non-Fourier heat conduction models, arXiv: cond-mat. 2212--12922v1 (2022).  
\bibitem{compte1997} A. Compte and R. Metzler, The generalized Cattaneo equation for the description of anomalous transport processes, J. Phys. A: Math. Gen. \textbf{30}, 7277 (1997).
\bibitem{awad2019} E. Awad, On the time--fractional Cattaneo equation of distributed order, Physica A \textbf{518}, 210 (2019).
\bibitem{fernandez} G. Fernandez--Anaya, F. J. Valdes--Parada, and J. Alvarez--Ramirez, On generalized fractional Cattaneo's equations, Physica A \textbf{390}, 4198 (2011).
\bibitem{qi} H. Qi and X. Jiang, Solutions of the space--time fractional Cattaneo equation, Physica A \textbf{390}, 1876 (2011).
\bibitem{liu} L. Liu, L. Zheng, and Y. Chen, Macroscopic and microscopic anomalous diffusion in comb model with fractional dual--phase--lag model, Appl. Math. Model. \textbf{62}, 629 (2018).
\bibitem{roscani} S. D. Roscani, J. Bollati, and D. A. Tarzia, A new mathematical formulation for a phase change problem with a memory flux, Chaos Solit. Fract. \textbf{116}, 340 (2018).
\bibitem{alegria} F. Alegria, V. Poblete, and J. C. Pozo, Nonlocal in--time telegraph equation and telegraph processes with random time, J. Differential Eqs. \textbf{347}, 310 (2023).
\bibitem{luchko} Y. Luchko, F. Mainardi, and Y. Povstenko, Propagation speed of the maximum of the fundamental solution to the fractional diffusion--wave equation, Comput. Math. Appl. \textbf{66}, 774 (2013).
\bibitem{awad2021} E. Awad, T. Sandev, R. Metzler, and A. Chechkin, From continuous-time random walks to the fractional Jeffreys equation: Solution and properties, Int. J. Mass Heat Transf. \textbf{181}, 121839 (2021).
\bibitem{hamada} Y. M. Hamada, Solution of a new model of fractional telegraph point reactor kinetics using differential transformation method, Appl. Math. Model. \textbf{78}, 297 (2020).
\bibitem{metzler1998} R. Metzler and T. F. Nonnenmacher, Fractional diffusion, waiting--time distributions, and Cattaneo--type equations, Phys. Rev. E \textbf{57}, 6409 (1998).
\bibitem{metzler1999} R. Metzler and A. Compte, Stochastic foundation of normal and anomalous Cattaneo-type transport, Physica A \textbf{268}, 454 (1999).
\bibitem{gorska2020} K. G\'orska, A. Horzela, E. K. Lenzi, G. Pagnini, and T. Sandev, Generalized Cattaneo (telegrapher's) equations in modeling anomalous diffusion phenomena, Phys. Rev. E \textbf{102}, 022128 (2020).
\bibitem{gorska2021} K. G\'orska, Integral decomposition for the solutions of the generalized Cattaneo equation, Phys. Rev. E {\bf 104}, 024113 (2021).
\bibitem{koszt2014} T. Koszto{\l}owicz, Cattaneo-type subdiffusion-reaction equation, Phys. Rev. E {\bf 90}, 042151 (2014).
\bibitem{vya} V. A. Vyawahare and P. S. V. Nataraj, Fractional-order modeling of neutron transport in a nuclear reactor, Appl. Math. Model. \textbf{37}, 9747 (2013).
\bibitem{nikan} O. Nikan, Z. Avazzadeh, and J. A. Tenreiro Machado, Numerical approach for modeling fractional heat conduction in porous medium with the generalized Cattaneo model, Appl. Math. Model. \textbf{100}, 107 (2021).
\bibitem{moza} M. Mozafarifard, D. Toghraie, and H. Sobhani, Numerical study of fast transient non-diffusive heat conduction in a porous medium composed of solid-glass spheres and air using fractional Cattaneo subdiffusion model, Int. Commun. Heat Mass Transf. \textbf{122}, 105192 (2021).
\bibitem{liu2017} Z. Liu, A. Cheng, and X. Li, A second order Crank–-Nicolson scheme for fractional Cattaneo equation based on new fractional derivative, Appl. Math. Comput. \textbf{311}, 361 (2017).
\bibitem{beghin} L. Beghin, R. Garra, F. Mainardi, and G. Pagnini, The tempered space-fractional Cattaneo equation, Probabilistic Engineering Mechanics \textbf{70}, 103374 (2022).
\bibitem{vieira} N. Vieira, M. Ferreira, and M. M. Rodrigues, Time-fractional telegraph equation with $\psi$--Hilfer derivatives, Chaos Solit. Fract. \textbf{162}, 112276 (2022). 
\bibitem{barbero} G. Barbero and J. Ross Macdonald, Transport process of ions in insulating media in the hyperbolic diffusion regime, Phys. Rev. E \textbf{81}, 051503 (2010). 
\bibitem{kostrobij} P. P. Kostrobij, I. I. Grygorchak, F. O. Ivaschyshyn, B. M. Markovych, O. V. Viznovych, and M. V. Tokarchuk, Mathematical modeling of subdiffusion impedance in multilayer nanostructures, Math. Model. Comput. \textbf{2}, 154 (2015).
\bibitem{kl2009} T. Koszto{\l}owicz and K. D. Lewandowska, Hyperbolic subdiffusive impedance, J. Phys. A {\bf 42}, 05500 (2009).
\bibitem{lk2008} K. D. Lewandowska and T. Koszto{\l}owicz, Appliaction of generalized Cattaneo equation to model subdiffusion impedance, Acta Phys. Polonica B {\bf 39}, 1211 (2008).
\bibitem{meroz2011} Y. Meroz, I. M. Sokolov, and J. Klafter, Unequal twins: probability distributions do not determine everything, Phys. Rev. Lett. {\bf 107}, 260601 (2011).
\bibitem{dybiec1} B. Dybiec and E. Gudowska-Nowak, Discriminating between normal and anomalous random walks, Phys. Rev. E {\bf 80}, 061122 (2009).
\bibitem{schilling} R. L. Schilling, R. Song, and Z. Vondra\v{c}ek, {\it Bernstein Functions. Theory and Applications} (De Gruyter, Berlin, 2010).
\bibitem{tkoszt2004} T. Koszto{\l}owicz, From the solutions of diffusion equation to the solutions of subdiffusive one, J. Phys. A: Math. Gen. {\bf 37}, 10779 (2004).
\bibitem{gorenflo} R. Gorenflo, A. A. Kilbas, F. Mainardi, and S. V. Rogosin, {\it Mittag-Leffler Functions, Related Topics and Applications} (Springer, Berlin, 2014), p.84.
\bibitem{tk2023} T. Koszto{\l}owicz, Subdiffusion with particle immobilization process described by a differential equation with Riemann--Liouville--type fractional time derivative, Phys. Rev. E {\bf 108}, 014132 (2023).
\bibitem{hughes} B. D. Hughes, {\it Random Walk and Random Environments. Vol. I, Random Walks} (Clarendon, Oxford, 1995).
\bibitem{masoliver} J. Masoliver and K. Lindenberg, Continuous time persistent random walk: a review and some generalizations, Eur. Phys. J. B \textbf{90}, 107 (2017).
\bibitem{masoliver2016} J. Masoliver, Fractional telegrapher’s equation from fractional persistent random walks, Phys. Rev. E {\bf 93}, 052107 (2016). 
\bibitem{montroll1965} E. W. Montroll and G. H. Weiss, Random walk on lattices. II, J. Math. Phys. {\bf 6}, 167 (1965).
\bibitem{kdfut} T. Koszto{\l}owicz and A. Dutkiewicz, in preparation.
\bibitem{sandev2017} T. Sandev, I. M. Sokolov, R. Metzler, and A. Chechkin, Beyond monofractional kinetics, Chaos Solit. Fract. \textbf{102}, 210 (2017).
\bibitem{sandev2015} T. Sandev, A. Chechkin, H. Kantz, and R. Metzler, Diffusion and Fokker--Planck--Smoluchowski equations with generalized memory kernel, Fract. Cal. Appl. Anal. \textbf{18}, 1006 (2015).
\bibitem{sandev2015a} T. Sandev, N. Korabel, H. Kantz, I. M. Sokolov, and R. Metzler, Distributed order diffusion equations and multifractality: Models and solutions, Phys. Rev. E \textbf{92}, 042117 (2015).
\bibitem{sandev2018} T. Sandev, R. Metzler, and A. Chechkin, From continuous time random walks to the generalized diffusion equation, Fract. Cal. Appl. Anal. \textbf{21}, 10 (2018).
\bibitem{sandev2019} T. Sandev, Z. Tomovski, J. L. A. Dubbeldam, and A. Chechkin, Generalized diffusion--wave equation with memory kernel, J. Phys. A: Math. Theor. \textbf{52}, 015201 (2019).
\bibitem{chechkin2021} A. Chechkin and I. M. Sokolov, Relation between generalized diffusion equations and subodrination schemes, Phys. Rev. E \textbf{103}, 032133 (2021).
\bibitem{aot} G. G. Anderson and G. A. O'Toole, Innate and induced resistance mechanisms of bacterial biofilms, Current Topics in Microbiology and Immunology \textbf{322}, 85 (2008). 
\bibitem{mot} T. F. C. Mah and G. A. O'Toole, Mechanisms of biofilm resistance to antimicrobial agents, Trends in Microbiology \textbf{9(1)}, 34 (2001). 


\end{thebibliography}
\end{document}